\documentclass[graybox, secnum]{svmult}

\usepackage{mathptmx}       
\usepackage{helvet}         
\usepackage{courier}        
\usepackage{type1cm}        
\usepackage{makeidx}         
\usepackage{graphicx}        
\usepackage{multicol}        
\usepackage[bottom]{footmisc}
\usepackage{hyperref}        
\usepackage{soul}            
\hypersetup{colorlinks=true,urlcolor=blue}
\usepackage[square,numbers]{natbib}
\makeindex             
\def\gsim{\mathrel{\rlap{\lower 4pt \hbox{\hskip 1pt $\sim$}}\raise 1pt
\hbox {$>$}}}
\def\lsim{\mathrel{\rlap{\lower 4pt \hbox{\hskip 1pt $\sim$}}\raise 1pt
\hbox {$<$}}}

\begin{document}
\title*{Stellar evolution, SN explosion, and nucleosynthesis}
\author{Keiichi Maeda\thanks{corresponding author}}
\institute{Keiichi Maeda \at Department of Astronomy, Kyoto University, Kitashirakawa-Oiwake-cho, Sakyo-ku, Kyoto 606-8502, Japan \email{keiichi.maeda@kusastro.kyoto-u.ac.jp}
}
%
%
\maketitle
\abstract{Massive stars evolve toward the catastrophic collapse of their innermost core, producing core-collapse supernova (SN) explosions as the end products. White dwarfs, formed through evolution of the less massive stars, also explode as thermonuclear SNe if certain conditions are met during the binary evolution. Inflating opportunities in transient observations now provide an abundance of data, with which we start addressing various unresolved problems in stellar evolution and SN explosion mechanisms. In this chapter, we overview the stellar evolution channels toward SNe, explosion mechanisms of different types, and explosive nucleosynthesis. We then summarize observational properties of SNe through which the natures of the progenitors and explosion mechanisms can be constrained. }

\vspace{1ex}

\section{Overview}

\begin{figure}[t]
\centering
\includegraphics[width=\columnwidth]{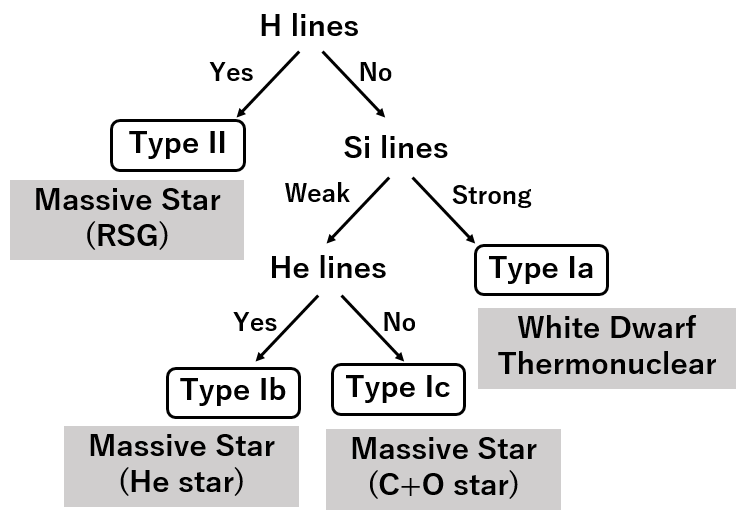}
\caption{Classification of SNe based on the properties of spectral-line features seen in their maximum-light spectra.
}
\label{fig:classification}
\end{figure}

Supernovae (SNe) announce catastrophic demise of stars. Their luminosities typically reach to $\sim 10^{43}$ erg s$^{-1}$ (with large variations), i.e., comparable to the typical luminosity of a galaxy in the optical wavelengths. The kinetic energy of the ejected materials, which is eventually inserted to its surroundings, is typically $\sim 10^{51}$ erg which is comparable to the whole energy budget of the Sun in its entire life of $\sim 10^{10}$ yr. With the rate of $\sim 0.01$ SN yr$^{-1}$, SNe provide the total energy of $\sim 10^{59}$ erg in the Hubble time into a galaxy they belong to, which is not negligible as compared to (or can even exceed) the binding energy of a galaxy. SNe thus play an important role in shaping the local, or even global, environment of a galaxy, e.g., by affecting the star-formation rate in the surrounding region. 

SNe are one of the most important events as the origin of heavy elements in the cosmic inventory. They not only eject heavy elements synthesized during the hydrostatic stellar evolution into the space, but create various elements, especially intermediate mass elements (IMEs) and iron-group elements (IGEs, or Fe-peak elements), at the moment of the explosion. SNe are thus key players in the Galactic chemical evolution. 

SNe show diverse observational properties, which reflect various pathways in the stellar evolution toward the end of their lives. To demonstrate this point, Fig \ref{fig:classification} shows a classical classification scheme based on spectral properties of SNe at their brightest (maximum-light) phase \cite{filippenko1997} (see Section 5 for details); this classical classification scheme relies mostly on the chemical composition in the outermost layer of the ejected material and thus in the progenitor stars; type Ia SNe are believed to be a thermonuclear explosion of a massive C+O white dwarf (WD), while the other classes shown in Fig. \ref{fig:classification} are all from the core-collapse of a massive star; type II SNe are an explosion of a Red-supergiant (RSG), type Ib and Ic SNe are an explosion of a (non-degenerate) He and C+O star, which are produced during the evolution of massive stars in which the outer H-rich envelope (or even He-rich layer) has been stripped away either by a strong stellar wind or binary interaction. There is a class of type IIb SNe (not shown in Fig. \ref{fig:classification}), which show transitions from type II to Ib in their spectral features; the SN IIb progenitors are believed to be similar to those of SNe Ib (i.e., a He star) but with a small amount of the H-rich envelope still attached at the time of the SN explosion. In summary, SNe are the end-products of both intermediate mass stars (after the formation of a WD) and massive stars (at the formation of a neutron star, NS, or even a black hole, BH). SNe thus provide an irreplaceable opportunity to study stellar evolution in the final phase, which is otherwise difficult by other means. 

In this chapter, we will first summarize basic principles of stellar evolution toward SN explosions, and provide basic pictures on the explosion mechanism(s); massive star evolution and core-collapse SNe in Section 2, and evolution of a WD in a binary and thermonuclear SNe in Section 3. Key concepts of explosive nucleosynthesis are introduced in Section 4. In Section 5, we will summarize emission processes of SNe, and discuss how observational features of SNe can be interpreted in terms of their progenitors and explosion mechanisms. We will also briefly mention mechanisms for high-energy and radio emissions from SNe. The review is closed in Section 6 with a summary.

\section{Massive Star Evolution and Core-Collapse Supernovae}

\subsection{Core Evolution Toward the Iron-Core Formation}

\begin{figure}[t]
\centering
\includegraphics[width=\columnwidth]{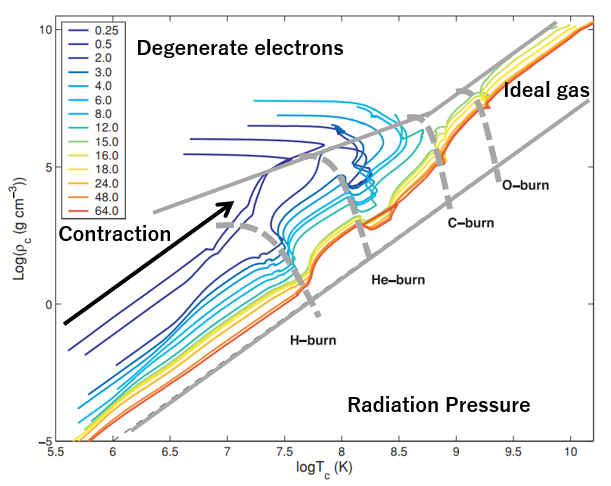}
\caption{The evolutionary pathways of stars with different masses in their central properties (Reproduction of \cite{kovets2009} with modifications). 
}
\label{fig:evolution}
\end{figure}

Theory of stellar evolution forms a basic framework for many branches of astronomy and astrophysics. It is a matured and classical field, but at the same time it continues to be in the forefront of astronomy with many outstanding problems yet to be solved. In this section, we provide some key (minimal) concepts that control the stellar evolution as basic rules, in an introductory and rather qualitative/simplified manner. Even within the standard framework of quasi-static evolution of a spherical star, it is already a highly non-linear problem, and many interesting and important phenomena and processes are not described by such a simplified treatment. We therefore recommend the readers to access classical textbooks with high reputation, such as \cite{clayton,kippenhahn}. There are also many modern text books at an introductory level, which may serve as the run-up before tackling to these standard textbooks. 

As a very rough, zeroth-order approximation, we may describe a star as a uniform sphere with the mass $M$ and radius $R$, with the sharp drop of the pressure at its surface. By integrating the hydrostatic balance ($dP/dr = -GM_r\rho / r^2$ together with $dM_r/dr = 4 \pi \rho r^2$, where $M_r$ is the enclosed mass below given radius $r$ within the star) for the whole star, it then requires the following; 
\begin{equation}
P \propto \frac{M^2}{R^4} \ , 
\end{equation}
where $P$ is the (central) pressure. Here, we are concerned with the scaling relation thus omit the coefficient (which is dependent on the detailed structure). It can also be viewed as the balance between the internal energy and the gravitation energy, i.e., $(4\pi R^3 / 3) P \propto (GM^2/R)$ which reduces to the same relation, noting the pressure is proportional to the internal energy. Assuming that the pressure is provided by the ideal gas ($P \propto \rho T$, with $\rho$ and $T$ the (central) density and temperature), we can derive the relation between the central density and temperature as 
\begin{equation}
\rho \propto T^3 M^{-2} \ \ \ {\rm or} \ \ \ T \propto \rho^{1/3} M^{2/3} \ .
\end{equation}
This relation has two important consequences: (1) For a given hydrostatic configuration (or roughy for given $T$ if the energy balance is maintained by the nuclear energy generation), a more massive star is less dense. (2) When the release of the gravitational binding energy is the only energy-generation process, the star collapses toward the higher density following the path in the $(T, \rho)$ plane as given by equation 2 (which is essentially parallel for stars with different masses). Not only the density but the temperature increase through the contraction, which is also viewed as a consequence of the Virial theorem for an object bound by the self-gravity. 

This behavior relies on the ideal gas Equation-of-State (EOS). Once the pressure is mainly provided by degenerate electrons, the binding energy released by the contraction is no more channeled into the increase in the thermal energy (i.e., no or little temperature increase anymore). The boundary can be estimated by equating the degenerate pressure at zero temperature ($P \propto \rho^{5/3}$ in the non-relativistic regime which is of main interest during the stellar evolution) and the thermal pressure ($P \propto \rho T$), yielding 
\begin{equation}
\rho \propto T^{3/2} \ \ \ {\rm or} \ \ \ T \propto \rho^{2/3} \ .
\end{equation}
Importantly, the line defining this boundary is flatter than the path of the contraction (eq. 2) in the $(T-\rho)$ plane, or steeper in the $(\rho-T)$ plane. 

The basic picture of stellar evolution obtained through detailed numerical calculations (an example shown in Fig. \ref{fig:evolution}) can be roughly understood by the general and simple rules as derived above. An opaque gas as born through contraction of the interstellar matter (ISM), as a seed of a star, keeps contracting toward the higher density and temperature. A more massive (proto-)star follows a track in the $(T-\rho)$ plane at a lower-density side. When the central temperature reaches to $\sim 10^{7}$ K, the hydrogen burning is initiated at the center, which balances the energy loss from the surface of the star; it is the main-sequence (MS) stage. 

Once hydrogen is (completely) consumed in the central region, the star is now described as a He core plus a H-rich envelope. The bottom of the envelope is still energized by the (shell) H-burning but the nuclear energy generation is now missing in the He core. The core then starts contracting, with a track in the $T-\rho$ plane again described by equation 2 (with $M$ and $R$ now replaced by the core properties, noting that the core mass is generally correlated with the zero-age main-sequence (ZAMS) mass ($M_{\rm ZAMS}$) and a more massive star/core always takes a lower-density track). The H-rich envelope reacts to the core contraction in a way that it expands to become a giant. Eventually the He-burning can be ignited, if the He core (therefore the initial mass) is sufficiently massive to avoid the dominance of the degenerate pressure before reaching to $\sim 10^{8}$ K. 

The subsequent evolution follows essentially the same way, i.e., repeated sequences of exhaustion of fuels in the present burning stage, core contraction, and then initiation of a new burning stage, toward more advanced burning stages ultimately toward the formation of an Fe core. The key that mainly determines the fate of a star is whether the electron degenerate pressure overwhelms the thermal pressure during the evolution, and if so, when it takes place. Given the lower-density track for a more massive star in the $(T-\rho)$ plane, a sufficiently massive star ends up with the formation of the Fe core. A less massive star will reach to the degenerate regime before the formation of the Fe core; for example, if the core becomes degenerate after the He-burning but before the C-burning, the core is essentially a C+O WD; this will become a C+O WD once the envelope is ejected to form a planetary nebula. 

Based on these considerations (and results of detailed stellar-evolution calculations), the end points of stellar evolution are summarized as follows (but note that the masses defining the boundaries can be dependent on the details, with different predictions from different simulations) \cite[e.g.,][]{nomoto1984,woosley1995,rauscher2002,heger2003,limongi2003,langer2012}: 
\begin{itemize}
\item {\bf $M_{\rm ZAMS} \lsim 8 M_\odot$: }  Formation of a WD. The dividing mass between a C+O WD and He WD is $M_{\rm ZAMS} \sim 0.5 M_\odot$. 
\item {\bf $8 M_\odot \lsim M_{\rm ZAMS} \lsim 10 M_\odot$: } The ONeMg core is formed but degenerates before entering into the next burning stages. The outcome will be either a ONeMg WD or an SN explosion following the ONeMg core collapse.
\item {\bf $M_{\rm ZAMS} \gsim 10 M_\odot$: } Formation of an Fe core, followed by a subsequent core collapse and an SN explosion. 
\end{itemize}

The time scale of the nuclear burning becomes shorter toward more advanced stages \cite[e.g.,][]{heger2003}. For example, for a star with $M_{\rm ZAMS} \sim 10 M_\odot$, the whole life time is $\sim 10^7$ yrs. The star stays in the H-burning MS stage for $\sim 90$\% of the whole life time, followed by the core He-burning stage lasting for $\sim 10^6$ yrs. The core C-burning is set in $\sim 1,000$ yrs before the core collapse. The O-burning lasts only for $\sim 1$ yr, and further burning stages have much shorter time scale. 

\subsection{Core-Collapse Supernova (CCSN) Explosion Mechanism}

\begin{figure}[t]
\centering
\includegraphics[width=0.45\columnwidth]{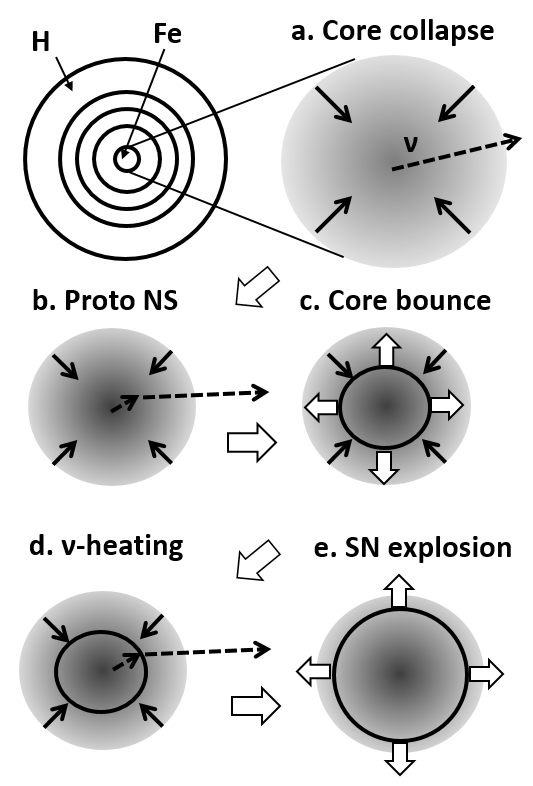}
\includegraphics[width=0.5\columnwidth]{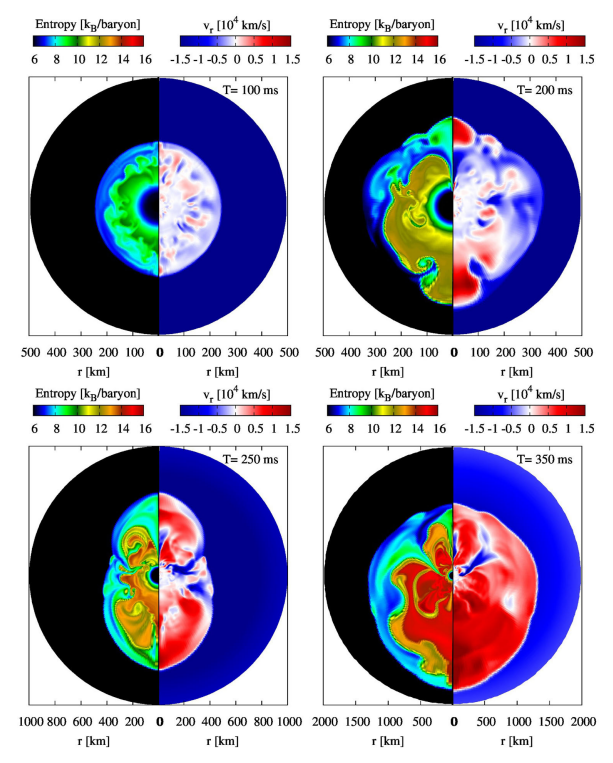}
\caption{Core-collapse Supernova (CCSN) explosion mechanism. The left panel shows a schematic picture. The right panel is an example of numerical simulations \cite{suwa2015}. 
}
\label{fig:ccsn}
\end{figure}

In this section, we overview the processes that are relevant to the Fe core collapse (Fig. \ref{fig:ccsn}). Studying the CCSN explosion mechanism is a rapidly evolving field tackled with state-of-the-art simulations for which readers can find good reviews and/or recent articles, such as \cite{fryer2004,mezzacappa2005, kotake2006,janka2012,bruenn2016,burrows2021}. 

In the advanced burning stages, the cooling via neutrinos becomes dominant, and the core evolution is accelerated. Once the Fe core is formed, the lack of the nuclear energy generation leads to the core collapse. Due to the increasing density and temperature, the core reaches to Nuclear Statistics Equilibrium (NSE; see Section 4 for details), through which Fe (or Ni) is photodisintegrated to nucleons and alpha particles: 
\begin{eqnarray}
^{54}_{26}{\rm Fe} + \gamma & \leftrightarrow & 13 ^{4}_{2} {\rm He} + 2 { \rm n} \ , \nonumber\\
^{56}_{26}{\rm Fe} + \gamma & \leftrightarrow & 13 ^{4}_{2} {\rm He} + 4 {\rm n} \ ,\\
^{4}_{2}{\rm He} + \gamma & \leftrightarrow & 2 {\rm p} + 2 {\rm n} \nonumber \ .
\end{eqnarray}
Through the photodisintegration, the thermal energy is further extracted from the core, accelerating the collapse. At the high density ($\gsim 10^{11}$ g cm$^{-3}$), electron captures also play an important role, due to the increasing electron Fermi energy that forbids the neutron decay; 
\begin{equation}
{\rm p} + {\rm e}^{-} \leftrightarrow {\rm n} + \nu_{\rm e} \ .
\end{equation}
This reduces the number of free electrons therefore the pressure, again accelerating the collapse. Indeed, for $M_{\rm ZAMS}$ in the range of $\sim 8-10 M_\odot$, the electron captures trigger the collapse of a degenerate ONeMg core. As a consequence of these processes, the core will evolve toward an increasingly dense and neutron-rich structure as the collapse proceeds. 

At some point, the density becomes sufficiently high so that the neutrinos are no more freely escaping from the core ($\gsim 10^{13}$ cm).  This proto-neutron star (NS) further collapses to reach to the nuclear density ($\sim 10^{14}$ g cm$^{-3}$), at which the collapse is suddenly stopped, forming the bounce shock on top of the proto-NS \cite{colgate1966}. The shock wave however finds it difficult to penetrate outward all the way against the ram pressure inserted by the infalling materials. In addition, the photodisintegration keeps extracting the energy from the shock wave. The shock is therefore stalled at $\sim 2 \times 10^{7}$ cm. Further energy deposition by neutrinos emitted from the newly-formed, hot proto-NS is believed to be a key to reviving the shock wave, which then leads to the SN explosion by expelling all the materials above the NS. This is the standard CCSN mechanism, called the delayed neutrino-heating mechanism \cite{bethe1985}. 

On the energetic ground, the energy source is the binding energy of the newly-formed NS; 
\begin{equation}
E_{\rm NS} \sim \frac{G M_{\rm NS}^2}{R_{\rm NS}} \sim 5 \times 10^{53} \ {\rm erg} \ ,
\end{equation}
if we use $M_{\rm NS} \sim 1.4 M_\odot$ and $R_{\rm NS} \sim 10^6$ cm for the mass and the radius of the newly-formed NS. Most of this energy is emitted by neutrinos as confirmed by the detection of neutrinos from SN 1987A \cite{bionta1987,hirata1987}. Compared to this energy budget is the kinetic energy of the ejected material by an SN explosion; this is typically an order of $\sim 10^{51}$ erg (see Section 5). Therefore, the conversion of $\sim 0.1 - 1$\% of the neutrino energy must be realized; this is a key problem in the CCSN explosion mechanism. It had been regarded to be a serious problem for several decades as virtually all the dedicated simulations (performed under the spherically-symmetric assumption) failed to revive the stalled shock \cite{liebendofer2001,sumiyoshi2005}. It is now believed that multi-dimensional effects are essentially important in the CCSN explosion mechanism, with an increasing number of reports for successful explosions in computer simulations. However, it is still a challenge to reproduce the explosions which look like what are observed as SNe; some key ingredients may still be missing, and the issue has been under further investigation intensively by various groups (see, e.g., the review articles introduced at the beginning of the section). 

\subsection{Core-Collapse Supernova Progenitors}

The evolution of the core as described in Section 2.1 represents only a single face of the stellar evolution. The envelope is the other important ingredient characterizing the nature of a star, evolution of which is generally decoupled from the core evolution in the advanced stage. The key process here is a mass loss, either through a strong stellar wind or binary mass transfer (which can be either stable or unstable, including a common envelope phase), or both. A good review for the roles of the mass loss processes and the relation to SN progenitors can be found, e.g., in \cite{langer2012}. 

\begin{figure}[t]
\centering
\includegraphics[width=0.8\columnwidth]{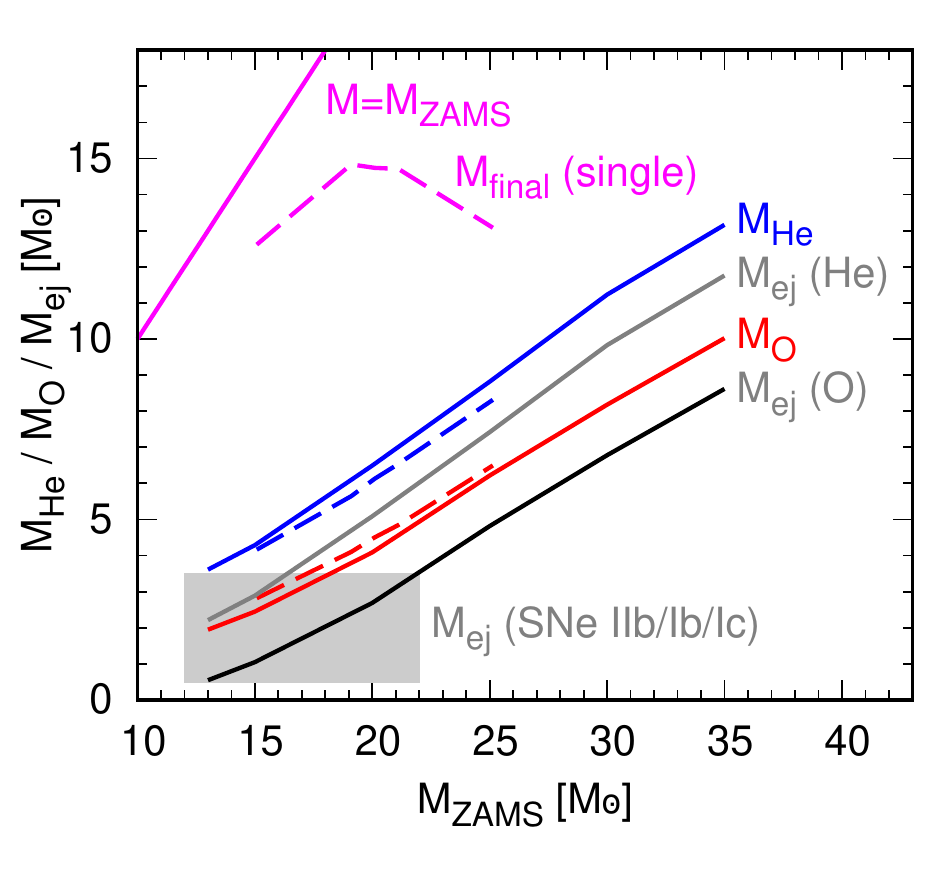}
\caption{The relations between the ZAMS mass and the core mass of an evolved star (i.e., SN progenitor), for the He core (blue) and the O core (red). Two (single) stellar models are shown; \cite{limongi2003} shown by the solids lines and \cite{rauscher2002} by the dashed lines. The final mass, with the mass loss given by a stellar wind (without a binary interaction), is also shown for the model by \cite{rauscher2002}. In addition, the expected ejecta masses are shown for bare He stars ($M_{\rm He} - M_{\rm NS}$; gray line) and for C+O stars ($M_{\rm O} - M_{\rm NS}$; black line), based on the models by \cite{limongi2003}, assuming $M_{\rm NS} = 1.4 M_\odot$. A range of the ejecta masses for SNe IIb/Ib/Ic (which are believed to be explosions of either a He star or C+O star for which all or most of the H-rich envelope has been ejected before the SN explosion), as inferred by observational data, is shown by the gray area \cite{lyman2016} (see Section 5 for details). 
}
\label{fig:mass}
\end{figure}

If most of the mass is retained until the He core is well developed (as a reasonable assumption in many channels), the He core mass is basically determined by the ZAMS mass since the track of the evolution of the core properties is mainly controlled by the ZAMS mass (Section 2) \cite[but see, e.g.,][]{laplace2021}. The same argument applies for the C+O core mass. Fig. \ref{fig:mass} shows these relations. The relation has an important implication for several classes of SNe, called SNe IIb, Ib, or Ic, whose progenitors are believed to be a (nearly) bare He or C+O star with a large fraction of the envelope stripped away during the evolution (Fig. \ref{fig:classification} and Sections 5). 


In most of the prescription used for the stellar-wind mass loss \cite{vink2001,vanloon2005}, the mass-loss rate is generally stronger for a more massive star. This is a reason why the single-star evolution model sequence by \cite{rauscher2002} (Fig. \ref{fig:mass}) has a peak in the final mass (i.e., the progenitor mass, including the H-rich envelope) around $\sim 20 M_\odot$. For sufficiently large $M_{\rm ZAMS}$, this effect alone can remove all the H-rich envelope, leading to the the formation of a bare He or even C+O star, i.e., a Wolf-Rayet (WR) star. The details depend on the mass-loss prescription, but this transition generally takes place at  $M_{\rm ZAMS} \sim 20 - 30 M_\odot$ in various models. We note that the strength of the stellar wind is also dependent on the metallicity; for lower metallicities, the amount of the stellar-wind mass loss decreases, and thus the ZAMS mass for the transition between an RSG and a WR moves upward \cite[e.g.,][]{heger2003}. 

\begin{figure}[t]
\centering
\includegraphics[width=\columnwidth]{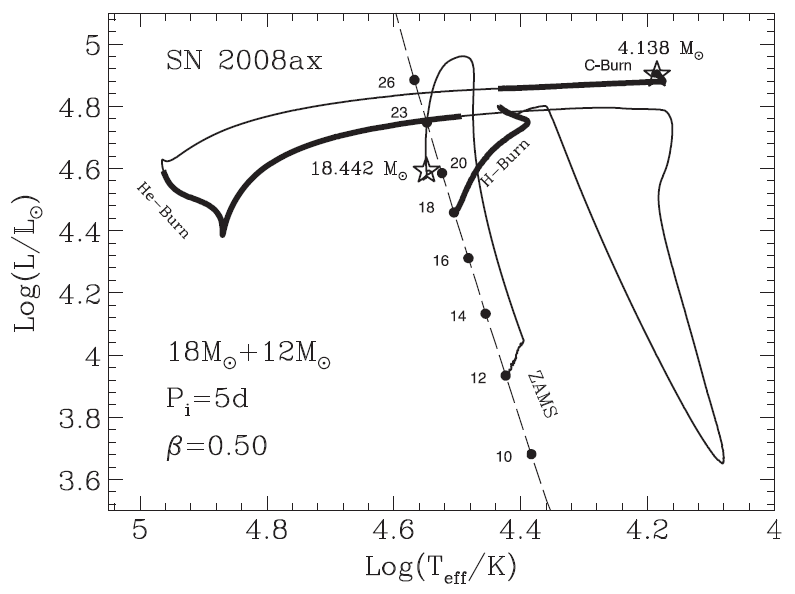}
\caption{An example of the binary evolution model \cite{folatelli2015}; the evolutionary tracks of the primary and companion stars are shown in the Hertzsprung-Russell diagram. 
}
\label{fig:binary_2b}
\end{figure}

The binary interaction scenario predicts different behaviors. It is indeed believed that a large fraction of massive star binaries experience a phase of strong binary interaction during their evolution toward the SN explosion \cite{sana2012}. Fig. \ref{fig:binary_2b} shows, as an example for a case of stable mass transfer, an evolution model for a binary system with the initial masses of $18 M_\odot$ and $12 M_\odot$. The initial orbital period is taken to be 5 days in this particular example. The primary starts experiencing the Roche-Lobe Over-Flow (RLOF) after the hydrogen core exhaustion. When the primary's mass decreases to $\sim 4.4 M_\odot$, the binary once detaches. The primary expands again after the helium core exhaustion, and undergoes the final RLOF; the primary's mass further decreases to $\sim 4.1 M_\odot$. The companion's mass increases through the mass accretion by the RLOF, reaching to the final mass of $\sim 18.4 M_\odot$ when the primary explodes as an SN. The SN progenitor in this model is essentially a He star but with a small amount of a H-rich envelope attached ($\sim 0.06 M_\odot$), which brings it into a blue-supergiant dimension ($\sim 40 R_\odot$ in its radius); it will thus explode as an SN IIb. In case the initial orbital period is shorter, then the primary will lose (nearly) an entire H-rich envelope and explode as an SN Ib. Further details on the binary evolution model toward SNe IIb/Ib/Ic (with stable mass transfer) can be found in, e.g., \cite{yoon2010,ouchi2017}

A common envelope also plays an important role, which can happen when the mass transfer rate is so high and the accreting star can not react to it in thermal time scale. For example, when a donor is substantially more massive than the accreting companion (i.e., an extreme mass ratio), the orbital shrinks as the mass transfer proceeds, leading to a rapid and unstable mass transfer and to the common envelope phase, where the companion star is engulfed by the primary's envelope. The details of the common envelope evolution has not yet been clarified despite intensive researches both in large simulations and observations \cite[e.g.,][and references therein]{iaconi2019}, and a phenomenological approach \cite{paczynski1976} is frequently (almost always) used to treat this phase in binary evolution models (including the so-called binary population models); this is based on the energy balance between the envelope's binding energy and (a fraction of) the (final) orbital energy. Namely, if there is a sufficient orbital energy exceeding the envelope's binding energy before the two cores marge, the outcome is a close binary followed by successful ejection of the primary's envelope (therefore the primary will become a He star for a massive-star binary, which is another possible progenitor channel for SNe Ib and Ic; \cite{nomoto1995}). Otherwise, the expected outcome is a merger of the cores of the two stars to become a (peculiar) single star (which can be a progenitor of some peculiar SNe, e.g., SN 1987A: \cite{morris2007,menon2017}). 

Yes another piece of potential importance is possible stellar activity toward the end of their lives. In the standard stellar evolution theory, it has been anticipated that the envelope reacts to the evolution of the core only slowly, despite the accelerated evolution of the core burning stages toward the end of the stellar life, i.e., the core-envelope decoupling. In the standard picture, the minimal response time can be estimated by the thermal time scale of the envelope, i.e., 
\begin{eqnarray}
t_{\rm th} & \sim & \frac{G M_{\rm c} M_{\rm env}}{R L}\\
& \sim & 70-700 \ {\rm yrs} \ \left(\frac{M_{\rm env}}{10 M_\odot}\right) \left(\frac{R}{10^{13} \ {\rm cm}}\right)^{-1} \ ,
\end{eqnarray}
where $M_{\rm c}$ is the core mass, $M_{\rm env}$ is the envelope mass, $R$ is the radius, and $L$ is the core luminosity. The range here covers the core luminosity being 10-100\% of the Eddington luminosity. The normalization here is given for an RSG progenitor. For a bare C+O or He star progenitor ($\sim$ a few $M_\odot$ and $\sim 10^{11}$ cm), the thermal time scale is even longer. Therefore, the thermal time scale is at least $\sim 100$ years. This means that the properties of the envelope are virtually `frozen' at least in the final century; the properties of the star in the final century are thus not expected to show either a trace of a rapid core evolution nor substantial variability. This simple picture, however, has been questioned in the last decade; there are mounting indications through observations of CCSNe that at least a fraction of SN progenitors show either the short-timescale variability or the rapid increase in the mass-loss rate in the final decades toward the SN. Section 5.3 will provide further details on this topical issue. 

\section{White Dwarfs in a Binary and Thermonuclear Supernovae}

\subsection{Thermonuclear Supernovae: Progenitors and Explosion Mechanisms}

Besides the CCSNe of a massive star, it has been widely accepted that Type Ia SNe (and some variants) are the outcome of a thermonuclear explosion of a C+O WD. Roughly speaking, a (massive) C+O WD can be disrupted by thermonuclear runway for the following reasons: (1) Due to the dominance of electron degenerate pressure, increase in temperature is not canceled by the cooling due to the expansion. The increasing reaction rate of carbon burning can thus lead to drastic and rapid increase in temperature, further accelerating the energy generation by the nuclear burning, i.e., the thermonuclear runaway. (2) The binding energy of a WD is at most comparable to the nuclear energy generation once the thermonuclear runaway incinerates a non-negligible fraction of the WD. As an example, for even a WD near the Chandrasekhar limiting mass, the binding energy is a few $\times 10^{50}$ erg. On the other hand, if $\sim 0.6 M_\odot$ of C+O materials are burnt to $^{56}$Ni (Section 4), the nuclear energy generation exceeds $10^{51}$ erg. In this section, we will first overview the possible conditions and mechanisms triggering the thermonuclear runaway within a C+O WD (Section 3.1), and then summarize possible evolutionary channels toward such situations (Section 3.2). 

\begin{figure}[t]
\centering
\includegraphics[width=0.8\columnwidth]{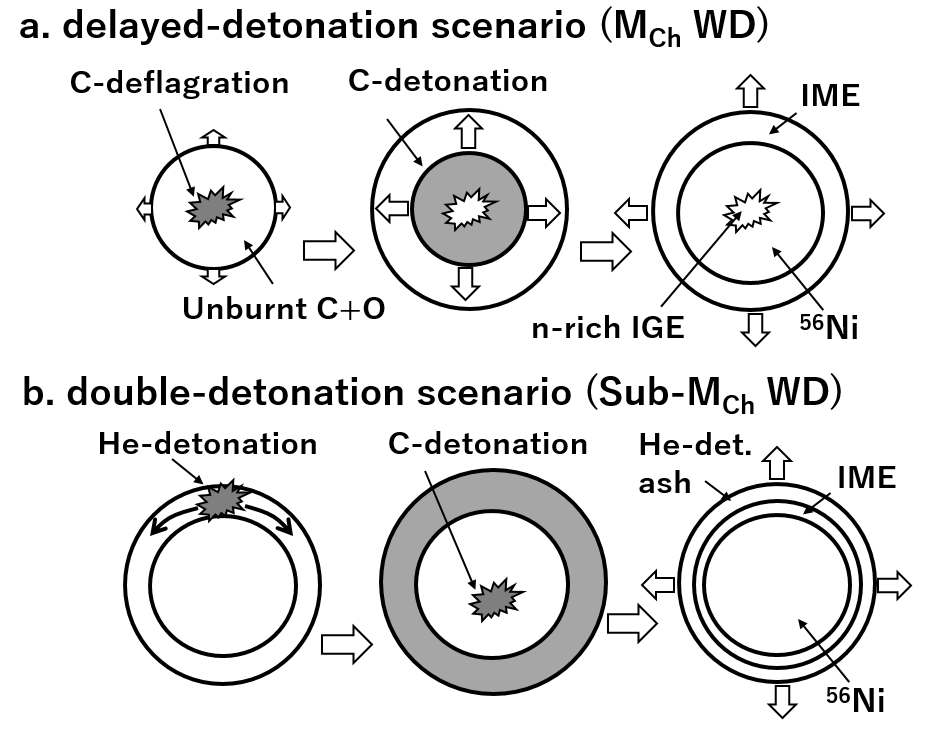}
\caption{Thermonuclear SN explosion mechanisms. 
}
\label{fig:sn1a}
\end{figure}

There are two popular scenarios for SN Ia progenitor(s) and explosion mechanism(s) (Fig. \ref{fig:sn1a}) \cite[see, e.g.,][for a review]{hillebrandt2000}; (1) the delayed-detonation mechanism on a C+O WD with nearly the Chandrasekhar-limiting mass ($M_{\rm Ch}$), and (2) double-detonation mechanism on a C+O WD mainly considered for a sub-Chandrasekhar-limiting mass (sub-$M_{\rm Ch}$) C+O WD. 

{\bf The delayed-detonation model on a $M_{\rm Ch}$ C+O WD:} As a C+O WD approaches to $M_{\rm Ch}$ in its mass, the central density and temperature increase in the highly-degenerate central region ($\gsim 10^{9}$ g cm$^{-3}$), to the point where the rate of nuclear energy generation exceeds the cooling rates by neutrinos and convection. It is then likely that the thermonuclear runaway starts near the center of the WD. 

There are two modes in the nature of the frame propagation, as in the chemical combustion in laboratory experiments. Deflagration is a sub-sonic frame in which the frame is driven by heat conduction or diffusion process. Detonation is a super-sonic frame in which the nuclear energy generation drives a shock wave, which keeps energized by the nuclear reactions behind it. Which mode is realized at the ignition is not completely clarified, but there are several arguments for the deflagration trigger \cite[e.g.,][]{nomoto1984,niemeyer1997,hillebrandt2000}: (1) The degenerate (Fermi) energy per volume is scaled as $\propto \rho^{4/3}$ in the relativistic regime, while the nuclear energy is so as $\propto \rho$. As such, the ratio of the thermal energy added by the nuclear reaction to the degenerate energy is scaled as $\propto \rho^{-1/3}$. Namely, excessive thermal energy to drive the pressure jump is less significant for higher density, preventing the formation of the detonation wave. (2) For typical conditions realized in the progenitor WD, the burning is likely in the flamelet regime at $\sim 10^{9}$ g cm$^{-3}$ (while in the distributed burning regime at $\sim 10^7$ g cm$^{-3}$). (3) Phenomenologically, prompt detonation deep inside a $M_{\rm Ch}$ WD produces too much $^{56}$Ni, which contradicts to the luminosities of individual SNe Ia and the Galactic chemical evolution. 

However, deflagaration alone is too weak to convert a good fraction of the C+O WD up to $^{56}$Ni and Fe-peak elements \cite{roepke2007}. The WD expands and the density decreases before the flame travels substantially, quenching the nuclear burning. The energy generation is thus limited, and it is possible that it would not disrupt a whole WD \cite[e.g.,][]{kromer2013}. In the delayed detonation model, it is hypothesized that the deflagration is tuned into detonation (`deflagration-to-detonation transition'; DDT) at the fuel density of $\sim 10^7$ g cm$^{-3}$ \cite{khokhlov1991,iwamoto1999}. The detonation then burns a large fraction of the C+O WD up to $^{56}$Ni. 

\begin{figure}[t]
\centering
\includegraphics[width=\columnwidth]{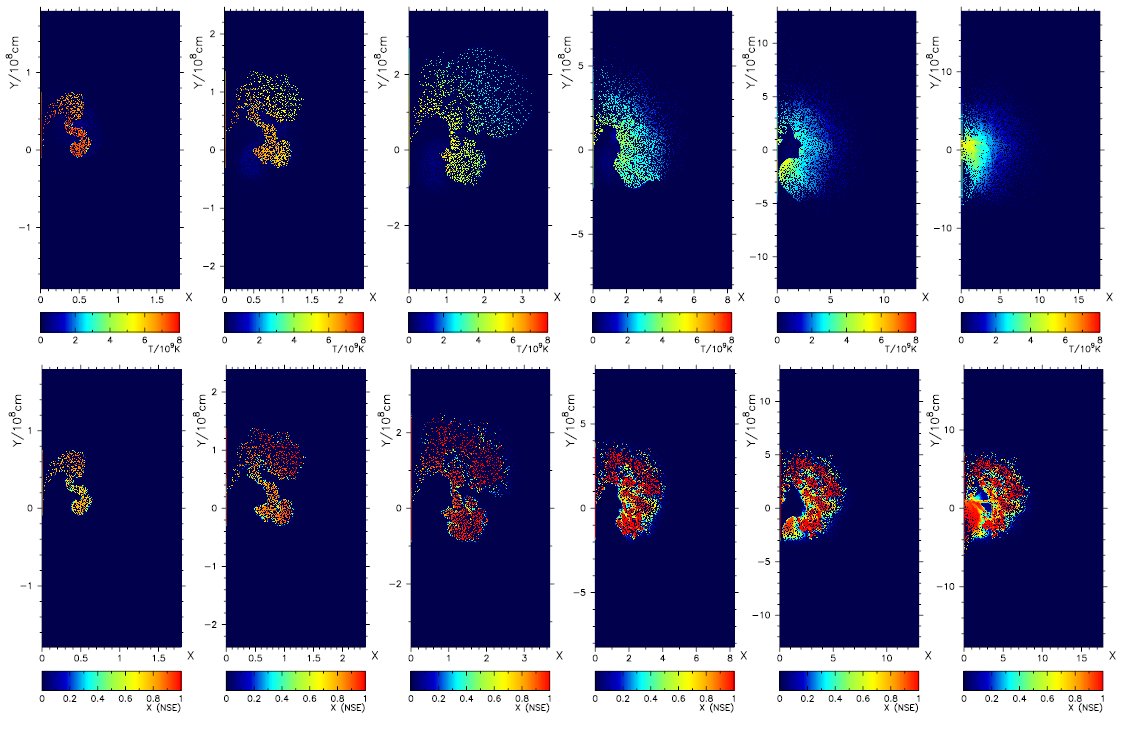}
\caption{An example of numerical (2D) simulations for the delayed detonation scenario \cite{maeda2010a}. The upper and lower panels show the temperature and the mass fraction of the Fe-peak elements, respectively. The temporal evolution is shown from the left to the right. 
}
\label{fig:ddt}
\end{figure}

{\bf The double-detonation model on a Sub-$M_{\rm Ch}$ C+O WD:} 
If the WD mass is substantially lower than $M_{\rm Ch}$, the central density and temperature are far below those required by a spontaneous ignition of carbon burning runaway. However, if there is an additional process that can dynamically raise the temperature there, thermonuclear runaway can take place. 

Helium accumulated on the WD surface can potentially play an important role. If a sufficiently large amount of He is accumulated on the WD surface, it can induce a detonation wave caused by unstable triple-alpha reactions \cite{bildsten2007}, as an analogy to a nova eruption due to accumulation of H-rich materials. The shock wave is inserted into the WD core. If the shock wave is sufficiently strong, it would initiate spontaneous carbon detonation near the center of the WD. This is the double-detonation model \cite{nomoto1982,livne1990,woosley1994,shen2009}. 

In both scenarios, multi-dimensional effects are critically important (see Fig. \ref{fig:ddt} as an example). In the delayed-detonation scenario, the deflagration phase is essentially asymmetric both in the trigger and flame propagation. The DDT then takes place at different spots, and the propagation of the detonation wave is affected by the asymmetrically distributed deflagration ash. In the double-detonation scenario, it is very likely that the He detonation is triggered on a spot rather than a shell. The subsequent He detonation, as well as the propagation of the shock wave into the C+O core, will be highly aspherical. Therefore, both scenarios have been intensively studied in terms of multi-dimensional simulations \cite[e.g.,][]{roepke2007b,kasen2009,maeda2010a,fink2010,seitenzahl2013,tanikawa2019,boos2021}. 

\subsection{Binary Evolution of a White Dwarf Toward Thermonuclear runaway}

One of the key issues related to the origin(s) of SNe Ia is which evolutionary channels lead to the formation of a (massive) C+O WD that satisfies conditions for thermonuclear runaway. In the delayed detonation mechanism, the main question is how a C+O WD reaches (nearly) to $M_{\rm Ch}$. In the double-detonation mechanism, the hurdle to create a massive C+O WD is lowered ($\sim 1 M_\odot$), but there is another requirement that a sufficiently massive He envelope is formed (or a sufficiently high temperature is reached) to the point that it undergoes unstable He burning. 

There are two major progenitor channels suggested for SNe Ia; single-degenerate (SD; \cite{whelan1973,nomoto1982}) and double-degenerate (DD; \cite{iben1984,webbink1984}) scenarios. The SD scenario considers a C+O WD as a mass accretor and a non-degenerate companion as a donor, while the DD scenario considers binary WDs including a merger of the two WDs. It should be emphasized that each explosion mechanism is not necessarily associated with each of the SD or DD channel; one may consider the delayed-detonation mechanism on a $M_{\rm Ch}$ C+O WD formed through a RLOF from a non-degenerate companion (SD) or as an outcome of merging WDs (DD). Similarly, the double-detonation mechanism may be realized by an accumulation of the He envelope through RLOF either from a non-degenerate companion (SD) or He-rich (or He) WD (DD), or even by a rapid accretion of He during a merging process of two WDs (DD). The SN Ia progenitor is a long-standing unresolved topic, for which a number of review articles can be found \cite[e.g.,][]{branch1995,wang2012,maoz2014,postnov2014,maeda2016,livio2018}. 

An example of the binary evolution channels (potentially) leading to SNe Ia is the following, where the directions of the evolution and mass transfer are indicated by arrows with different symbols (`$\Rightarrow$' for the former and `$\rightarrow$' for the latter) . Note that it is for a demonstration purpose, and there are many variants in details (see the review articles mentioned above). 
\begin{itemize}
\item[{\bf (a)}] \ {\bf MS + MS $\Rightarrow$ RG $\rightarrow$ MS $\Rightarrow$ CE $\Rightarrow$ C+O WD + MS:}\\ As the primary evolves and expands, it starts filling the RL (Roche Lobe). As the primary is more massive than the secondary, a likely outcome is an unstable transfer leading to a common envelope (CE). Unless the CE leads to a core merger, it will leave a close binary of a C+O WD and a MS. 
\item[{\bf (b)}] \ {\bf C+O WD $\leftarrow$ MS/RG $\Rightarrow$ SN (SD):}\\
As the non-degenerate secondary evolves, it starts filling the RL. The most plausible companion star in the SD scenario is either a MS ($\sim 2 M_\odot$) or an RG ($\sim 1 M_\odot$). The outcome of the RLOF depends on the mass-transfer rate ($\dot M$): 
  \begin{itemize}
  \item {\bf $\dot M \sim 10^{-7} -10^{-6} M_\odot$ yr$^{-1}$:} 
  The accretted materials are burnt to He and then to C in a steady-state manner. The WD mass thus increases, and can reach to $M_{\rm Ch}$ if the sufficient mass is provided by the donor ($\Rightarrow$ a $M_{\rm Ch}$ WD explosion in the SD; e.g., \cite{nomoto1984}). 
  \item {\bf $\dot M \gsim 10^{-6} M_\odot$ yr$^{-1}$:} Because of the rapid mass transfer, the WD cannot stably accrete the transferred mass in thermal time scale. The WD thus behaves like a giant with its envelope (formed by the accreted materials) inflating in its radius. The system thus undergoes a CE (then evolves to the point c below). However, it is also possible that the excessive mass in the accreted materials could be efficiently blown away from the system by a WD wind, with the effective accretion rate regulated to the steady-state limit. If this is realized, the system can lead to the thermonuclear explosion ($\Rightarrow$ a $M_{\rm Ch}$ WD explosion in the SD; e.g., \cite{hachisu1996}). 
  \item {\bf $\dot M \lsim 10^{-7} M_\odot$ yr$^{-1}$:} 
  The accreted materials are efficiently cooled and accumulated on the WD. When a critical mass is reached (which is lower for a more massive WD), the H-rich, accreted material undergoes thermonuclear runaway on the WD surface, i.e., a nova. It has not been clarified whether the WD mass indeed increases through the repeated accretion-nova evolution; it may become an SNe Ia ($\Rightarrow$ a $M_{\rm Ch}$ WD explosion in the SD; e.g., \cite{hachisu2001}) or not. 
  \end{itemize}
\item[{\bf (c)}] \ {\bf C+O WD $\leftarrow$ RG $\Rightarrow$ CE:}\\
Following the expansion of the non-degenerate secondary, the systems evolves into the CE phase. Further evolution may branch into different outcomes depending on the result of the CE interaction and the nature of the secondary. 
\item[{\bf (c1)}] \ \ \ {\bf CE $\Rightarrow$ Core merger $\Rightarrow$ peculiar SN Ia?}\\
The core of the secondary may merge into the primary (C+O) WD. If the secondary's core is a degenerate C+O core, it is essentially merging two C+O WDs as is similar to the DD system. If the secondary's core is either a degenerate or non-degenerate He core, the system may suffer from a He-ignited double-detonation explosion. These channels are thus potential pathways to peculiar SNe Ia surrounded by a massive CSM (core-degenerate scenario or its variants; e.g., \cite{sparks1974,soker2015,jerkstrand2020}). 
\item[{\bf (c2)}] \ \ \ {\bf CE $\Rightarrow$ C+O WD + C+O WD $\Rightarrow$ SN (DD)}\\
The CE leaves a compact binary system of two C+O WDs. Following the angular momentum loss by gravational waves, the two WDs may merge. If the sum of the two WD masses is far beyond $M_{\rm Ch}$, a prompt carbon detonation may disrupt both WDs, which may lead to a peculiar SN Ia ($\Rightarrow$ a sub-$M_{\rm Ch}$ explosion in the DD, noting that it is essentially `$M_{\rm Ch}$' in terms of the explosion condition; e.g., \cite{pakmor2010}). Otherwise, the outcome of the merger will be a massive WD surrounded by a hot envelope (formed by debris of the disrupted secondary). If the total mass is below $M_{\rm Ch}$, the end product is likely a single massive WD. If the mass is above $M_{\rm Ch}$, the WD core may eventually reach nearly to $M_{\rm Ch}$ by further accreting the C+O material from the envelope. If this is realized, it may explode as an SN Ia (a $M_{\rm Ch}$ explosion in the DD; \cite{iben1984,webbink1984}). However, the C+O core may be converted to a ONeMg core by off-center carbon deflagration during the envelope accretion; in this case, the end product is a $M_{\rm Ch}$ ONeMg WD, which may collapse to a NS through electron capture reactions \citep{saio1985,schwab2021}.  
\item[{\bf (c3)}] \ \ \ {\bf CE $\Rightarrow$ C+O WD + He WD $\Rightarrow$ SN (DD)}\\
In case the CE leaves a compact binary of a C+O WD and a He WD (or a C+O WD with a massive He envelope), further orbital shrink may lead to the He acretion to the primary WD, i.e., a potential AM CVn system \cite[e,g,][]{bildsten2007}. The outcome of the mass transfer is anologous to the SD case (point b above). For the high mass-transfer case, two WDs may merge, and the double detonation may result by the rapid He accretion during the merger process ($\Rightarrow$ a sub-$M_{\rm Ch}$ explosion in the DD; e.g., \cite{pakmor2013,shen2018,tanikawa2019}). For a moderate transfer rate, the He accretion may become steady-state, leading to formation of a $M_{\rm Ch}$ C+O WD ($\Rightarrow$ a $M_{\rm Ch}$ explosion in the DD; e.g., \cite{wang2009}). For the low mass-transfer case, the He envelope mass may reach to the condition for the He detonation ($\Rightarrow$ a sub-$M_{\rm Ch}$ explosion in the DD; e.g., \cite{shen2009}). 
\item[{\bf (c4)}] \ \ \ {\bf CE $\Rightarrow$ C+O WD + He star   $\Rightarrow$ SN (SD)}\\
This is similar to the case c3. As the non-degenerate He companion evolves and expands, it will start experiencing the RLOF to the primary C+O WD. For the moderate mass-transfer rate, the primary WD may burn He steadily on its surface and reach to $M_{\rm Ch}$ ($\Rightarrow$ a $M_{\rm Ch}$ explosion in the SD). For a low mass-transfer rate, the surface He detonation may result ($\Rightarrow$ a sub-$M_{\rm Ch}$ explosion in the SD). For the high mass transfer, the primary may reach to $M_{\rm Ch}$ if the WD wind stabilizes the mass accretion ($\Rightarrow$ a $M_{\rm Ch}$ explosion in the SD), or otherwise the system may undergo the merger, which is analogous to the point c1 above ($\Rightarrow$ a sub-$M_{\rm Ch}$ explosion or a peculiar SN). 
\end{itemize}

\section{Explosive Nucleosynthesis}

\begin{figure}[t]
\centering
\includegraphics[width=\columnwidth]{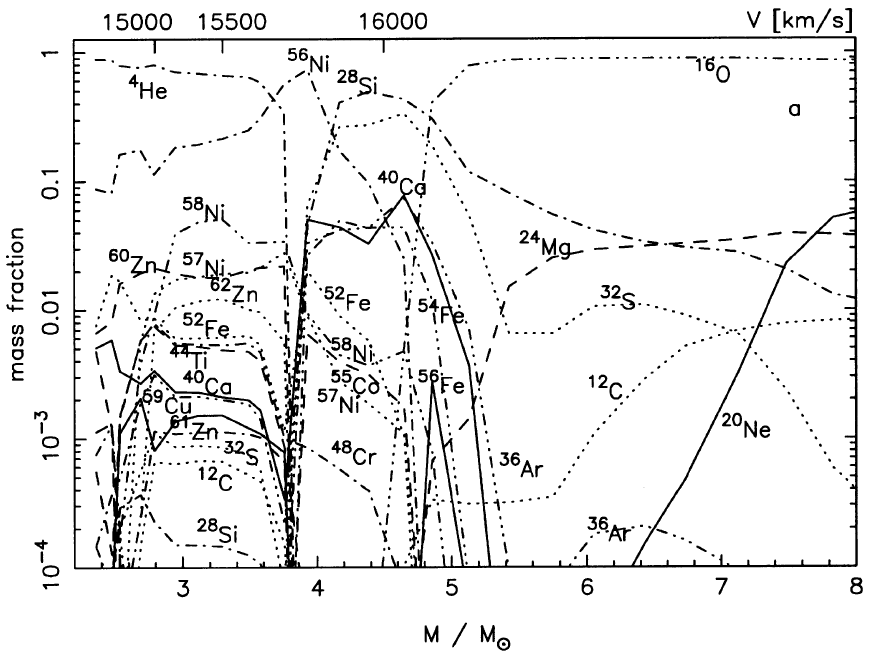}
\caption{An example of composition structure in the SN ejecta as an outcome of explosive nucleosynthesis in a CCSN \citep{maeda2002}. 
}
\label{fig:nucleosynthesis}
\end{figure}

The initial configuration at the onset of the explosion is either an onion-like structure for CCSNe (with the Si layer at the base, as the Fe core has collapsed to become a compact object) or the C+O composition for SNe Ia. The shock wave or thermonuclear flame inserted by the explosion mechanism then induces the so-called explosive nucleosynthesis, following its propagation toward the outer layer of the progenitor star. The SN ejecta thus have a layered composition structure as an outcome of the explosive nucleosynthesis, for which an example is shown in Fig. \ref{fig:nucleosynthesis}. We refer to  \cite{arnett,woosley1995,thielemann1996} that provide valuable basis on this topic. 

To the zeroth-order approximation, we may assume that the region behind the shock wave forms a fireball dominated by the radiation pressure, with uniform temperature. The temperature ($T$) behind the shock wave as a function of its radial location ($R$) can thus be estimated as follows \cite[e.g.,][]{maeda2009}; 
\begin{equation}
T  \sim  \left(\frac{3 E}{4 \pi a}\right)^{1/4} R^{-3/4} 
 \sim  1.3 \times 10^{10} \ {\rm K} \ 
 \left(\frac{E}{10^{51} \ {\rm erg}}\right)^{1/4}
 \left(\frac{R}{10^8 \ {\rm cm}}\right)^{-3/4} \ ,
\end{equation}
where $E$ is the explosion energy (which can be assumed to be roughly constant during the shock propagation for CCSNe, unless the explosion mechanism would be extremely `slow'; \citep{sawada2019}), and $a$ is the radiation constant. The temperature thus decreases as a function of time, following the expansion of the shock wave toward the outer region. We thus see that the temperature inside the shock is higher than $\sim 10^{9}$ K until the shock reaches to $R \sim 3 \times 10^9$ cm. Depending on the ZAMS mass, this can be comparable to the size of the C+O core. We therefore conclude that the explosive nucleosynthesis changes the composition in the substantial amount of materials in the innermost ejecta. 

The outcome of the explosive nucleosunthesis is roughly characterized by the peak temperature to which the material is heated behind the shock wave ($T_9 \equiv T/10^9$ K) \cite[e.g.,][]{arnett}. 
\begin{itemize}
\item {\bf $T_9 \gsim 5$ (Complete Si burning):} At the high temperature, both forward and reverse reactions in main reaction channels through the strong force proceed much faster than the typical evolution time scale with which the thermal condition changes (i.e., expansion and cooling). The compositions thus basically follow the Nuclear Statistical Equilibrium (NSE). Following the rapid temperature decrease, the reaction rates are becoming smaller and the abundance pattern `freezes out', leaving the NSE abundance pattern at the high temperature as the final outcome. In a typical condition realized in CCSNe, weak interactions are much slower, and thus the electron fraction ($Y_{\rm e}$, i.e., the number of protons divided by that of nucleons) is conserved. For the materials with the (roughly) equal numbers of protons and neutrons ($Y_{\rm e} \sim 0.5$, which applies to most part of the progenitor), the most abundant isotope is $^{56}$Ni (due to the combination of its being at the peak in the binding energy and the equal numbers of protons and neutrons in it), followed by He and other Fe-peak elements. 
\item {\bf $T_9 \sim 4-5$ (Incomplete Si burning):} The relative abundances within Fe-peak elements, as well as those within intermediate-mass elements (IMEs), roughly follow the NSE abundance pattern within each cluster, but the ratio between the amounts of the Fe-peak elements and the IMEs is not necessarily described by the NSE; it is called Quasi-NSE (QSE). This is (to the zeroth-order approximation) due to the bottleneck at $Z=20$, i.e., Ca having a high stability with a magic proton number. The outcome is Si, S, $^{56}$Ni, Ar, Ca, and Fe-peak elements, roughly in the decreasing order in the final mass fractions. 
\item {\bf $T_9 \sim 3-4$ (Oxygen burning):} The IMEs follow QSE, but the production of heavier elements is blocked by the Ca bottleneck. The abundant elements are O, Si, S, Ar, and Ca. 
\item {\bf $T_9 \sim 2-3$ (Carbon and neon burning):} In this temperature range, the explosive carbon and/or neon burning produces O, Mg, Si, and Ne. 
\end{itemize}

There are a few other processes especially in the high-temperature and/or high-density regime that have to be taken into account, depending on the situation: 
\begin{itemize}
\item {\bf $\alpha$-rich freeze out:} The outcome of the Si-burning regime depends on the condition for the freeze out. Either for higher peak temperature or lower density, the triple $\alpha$ reaction freezes out earlier, leaving a larger amount of He. This enhances the $\alpha$-capture reactions onto Fe-peak elements. As a result, the abundance pattern of Fe-peak elements is shifted, leaving a non-negligible amount of some specific isotopes that are difficult to produce in the usual Si burning. This includes the production of $^{44}$Ti and $^{48}$Ti (as a consequence of the chain of $^{40}{\rm Ca} \rightarrow ^{44}{\rm Ti} \rightarrow ^{48}{\rm Cr}$, which then decays to $^{48}$Ti), and $^{64}$Zn (through $^{56}{\rm Ni} \rightarrow ^{60}{\rm Zn} \rightarrow ^{64}{\rm Ge}$, which then decays to $^{64}$Zn), and so on. This process is especially important in the innermost region of the CCSN ejecta. 
\item {\bf $\nu$-driven wind in CCSNe:}  In the standard picture, the CCSN explosion mechanism is driven by a hot bubble above the proto-NS that is continuously heated by neutrinos (Section 2). This can also be regarded as a NS wind, which experiences distinctly higher temperature/entropy than the shock-heated material as discussed so far. It has been proposed as a candidate site for r-process nucleosynthesis \cite[e.g.,][]{qian1996}, but recent state-of-the-art CCSN simulations show that the wind more likely becomes proton-rich and thus the r-process nucleosynthesis is limited \cite[e.g.,][]{wanajo2018}. The nucleosynthesis outcome is therefore characterized by proton-rich Fe-peak elements/isotopes. 
\item {\bf $\nu$-process:} As the neutrinos emitted from the proto-NS pass through the region that has experienced the Si-burning, the neutrinos can interact with Fe-peak elements through weak interactions. This can be an important process especially in the production of relatively minor isotopes. Examples include $^{56}$Ni($\nu$, $\nu'$p)$^{55}$Co, which eventually decays to $^{55}$Fe and then to $^{55}$Mn \cite[e.g.,][]{yoshida2008}. 
\item{\bf electron capture:} The electron capture is important at high density. This is indeed one of the key processes in the formation of a proto-NS following a massive star collapse (Section 2). While a large fraction of the material which has experienced the electron capture reactions is `locked' onto a NS, a small fraction of materials may participate in the explosive nucloesynthesis under the neutron-rich condition (which, however, will be altered to become proton-rich in recent CCSM simulations; see above). Indeed, electron capture is even more important in determining the ejecta composition in SNe Ia. If the progenitor WD is sufficiently massive (i.e., close to the Chandrasekhar mass), the central density reaches to $\sim 10^9$ g cm$^{-3}$ at which the electron capture reactions leads to a neutron-rich condition (i.e., lower $Y_{\rm e}$). Unlike the CCSNe, this central region is ejected. Therefore, the innermost ejecta of SNe Ia are expected to have a large amount of neutron-rich Fe-peak elements such as $^{58}$Ni, if the progenitor WD is sufficiently massive; this has been proposed as one powerful diagnostics to discriminate the $M_{\rm Ch}$ WD and sub-$M_{\rm Ch}$ WD progenitor scenarios \cite[e.g.,][]{maeda2010b,yamaguchi2015}.
\end{itemize}

\begin{figure}[t]
\centering
\includegraphics[width=0.49\columnwidth]{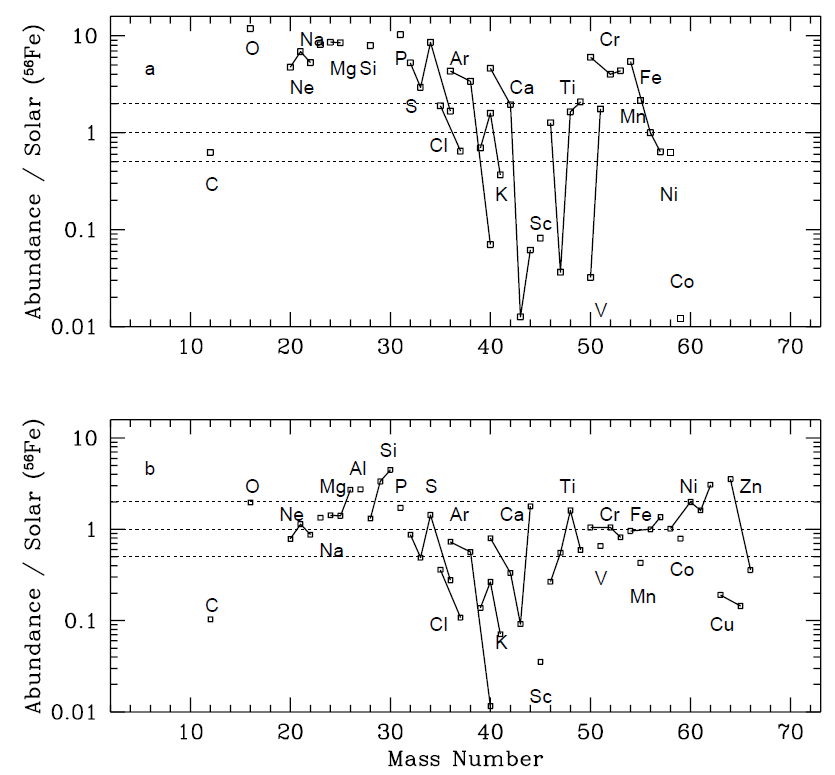}
\includegraphics[width=0.49\columnwidth]{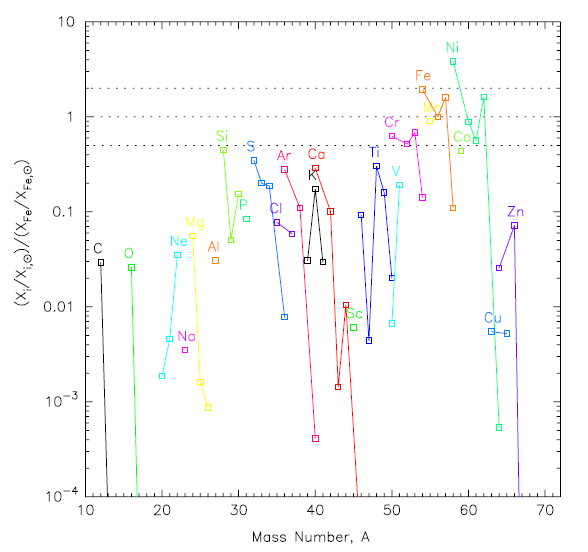}
\caption{Examples of nucleosynthesis yields for CCSNe (left panels; \cite{maeda2003}) and SNe Ia (right panel; \cite{maeda2010a}). In the left panel, the progenitor is a $16 M_\odot$ He star ($M_{\rm ZAMS} = 40 M_\odot$), and the two panels are for models with different masses of ejected $^{56}$Ni (larger for the lower panel; treated as the mass-cut; see the text). The right panel is for the W7 model \cite{nomoto1984b}, which belongs to the $M_{\rm Ch}$ WD explosion scenario. 
}
\label{fig:yields}
\end{figure}

The example shown in Fig. \ref{fig:nucleosynthesis} (for a CCSN) can be qualitatively understood by the above summary. The innermost region with abundant He followed by $^{56}$Ni is the region processes by the $\alpha$-rich freeze out in the complete Si-burning regime. Toward the outer region, the most abundant isotopes change from $^{56}$Ni to $^{28}$Si, and then to $^{16}$O; these regions have experienced the complete Si-burning (without $\alpha$-rich freeze out), the incomplete Si-burning, and then the O-burning (with ineffective explosive nucleosynthesis toward the surface). 

Examples of the nucloeynthesis yields, as integrated over all the ejected materials, are shown in Fig. \ref{fig:yields}. It is seen that generally CCSNe are characterized by a large mount of IMEs (produced mainly through the hydrostatic evolution), while SNe Ia are so by Fe-peak elements (through the explosive burning). For CCSNe, the model in this figure is parameterized by the so-called mass-cut, which determines the boundary between the NS/BH and the ejected materials; this has a direct link to the explosion mechanism (i.e., providing constraint on the explosion mechanism through the nucleosynthesis products; e.g., \cite{sato2021}). In the SN Ia model presented here, a large amount of stable Ni ($^{58}$Ni) is discerned, which is characteristic of the $M_{\rm Ch}$ WD scenario (see above).

\section{Emissions from Supernovae}

\subsection{Characteristic behaviors}

\begin{figure}[t]
\centering
\includegraphics[width=\columnwidth]{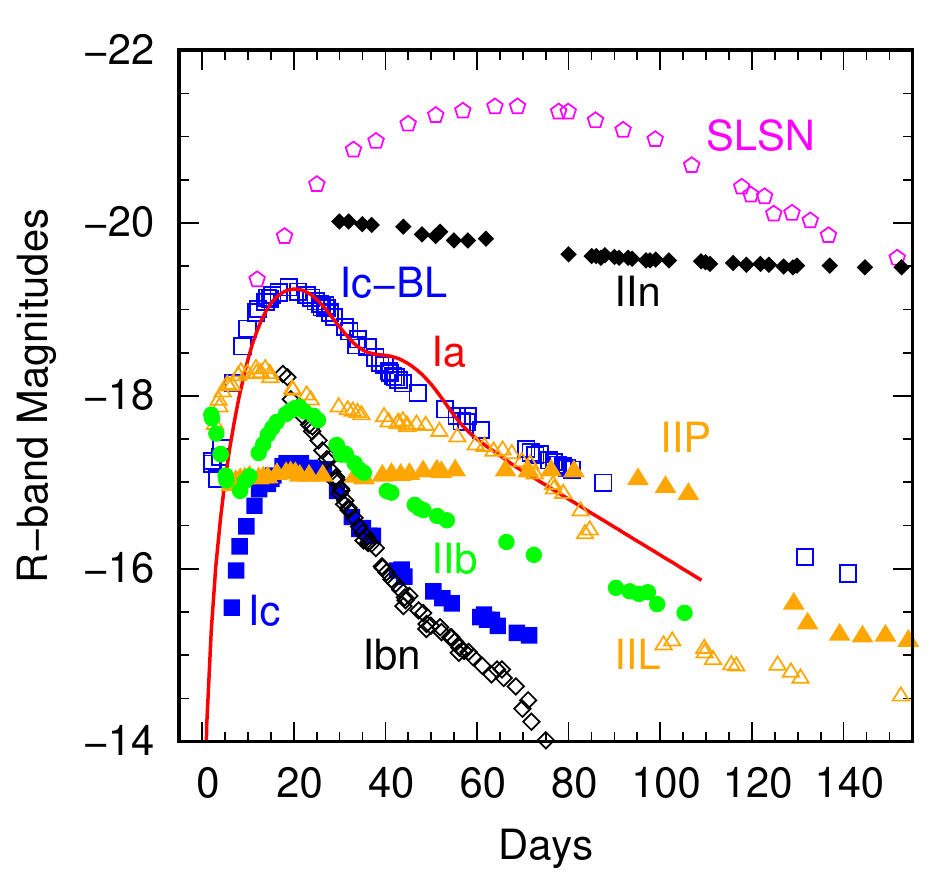}
\caption{Examples of optical ($R$-band) light curves of SNe of different types. It is the compilation of the data from the following sources: \cite{richmond1994,leonard2002,hsiao2007,pastorello2007,smith2007,valenti2008,clocchiatti2011,zhang2012,valenti2015}. 
}
\label{fig:LC}
\end{figure}

\begin{figure}[t]
\centering
\includegraphics[width=0.8\columnwidth]{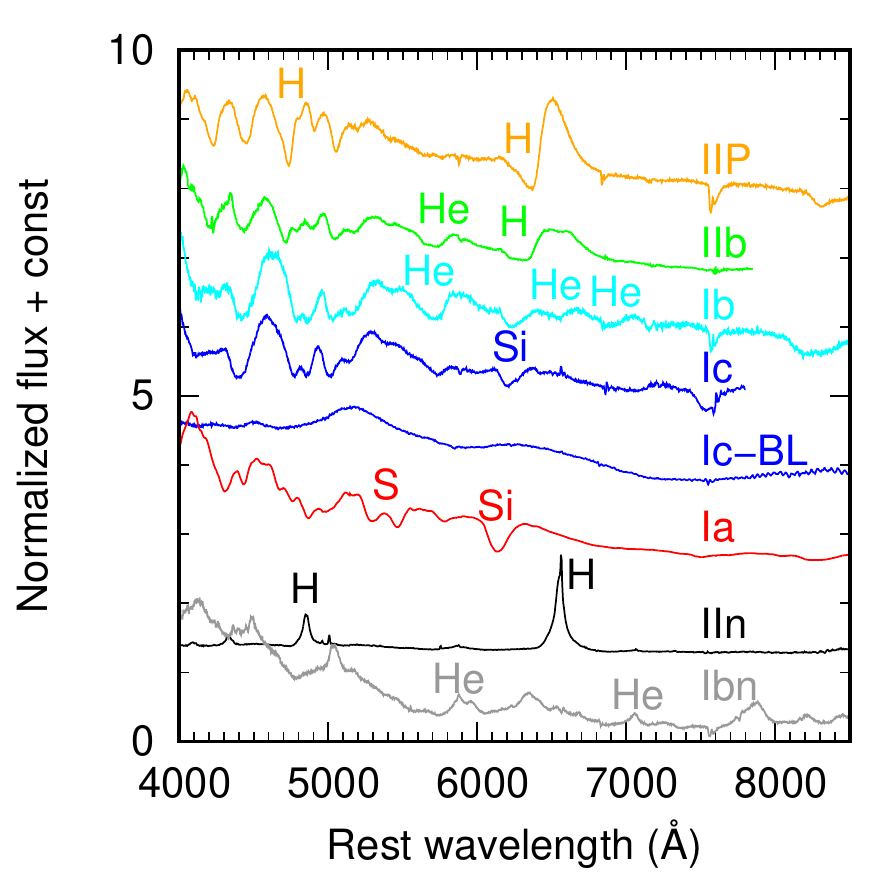}
\caption{Examples of (near) maximum-light spectra of SNe of different types. It is the compilation of the data from the following sources: \cite{barbon1995,hamuy2001,patat2001,valenti2008,anupama2009,zhang2012,pereira2013,srivastav2014}. The data are obtained from WISeREP (https://www.wiserep.org) \cite{wiserep}.
}
\label{fig:spec}
\end{figure}

The progenitor evolution, SN explosion mechanism and nucleosynthesis products manifest themselves in the observational data of SNe. Studying the natures of SN remnants (mostly in the Milky Way but also in the local group) and extragalactic SNe provides complementary approaches. In this contribution, we focus on extragalactic SNe. As excellent textbooks on the emission processes and observational properties of SNe, we refer the readers to \cite{arnett,branch}

Figs. \ref{fig:LC} and \ref{fig:spec} show typical light curves (LCs) and maximum-phase spectra for various types of SNe. In this chapter, we will overview how the natures of the progenitors and explosion mechanisms can be inferred and constrained based on these observational data.

Once a progenitor star is fully disrupted by an underlying explosion mechanism, the resulting SN is regarded as metal-rich ejecta expanding into the ISM or circumstellar matter (CSM). The characteristic velocity, $V$, is roughly given as follows; \begin{equation}
V \sim \sqrt{\frac{E_{\rm K}}{M_{\rm ej}}} \sim  7,000 \ {\rm km s}^{-1} \left(\frac{E_{\rm K}}{10^{51} \ {\rm erg}}\right)^{1/2} \left(\frac{M_{\rm ej}}{M_\odot}\right)^{-1/2}\ ,
\end{equation}
where $E_{\rm K}$ is the kinetic energy associated with the expansion of the ejecta, and $M_{\rm ej}$ is the ejecta mass. The diffusion time scale is then given as follows \cite[e.g.,][]{arnett};  
\begin{equation}
t_{\rm dif} \sim \frac{\kappa}{M_{\rm ej}}{\beta c R} \ , 
\end{equation}
where $\beta$ is a scaling constant that depends on the distribution of the opacity ($\kappa$) and density. The characteristic radial extent of the ejecta is simply described as $R = V t$, where $t$ is the time since the explosion (assuming $V t$ is substantially larger the progenitor radius, $R_0$). With $\beta \sim 13.8$ which is applicable to a range of the density/opacity distribution, the diffusion time is described as follows; 
\begin{equation}
t_{\rm dif} \sim 180 \ {\rm days} \left(\frac{\kappa}{0.2 \ {\rm cm}^2 \ {\rm g}^{-1}}\right) \left(\frac{M_{\rm ej}}{M_\odot}\right)^{3/2}  \left(\frac{E}{10^{51} \ {\rm erg}}\right)^{-1/2} \left(\frac{t}{\rm day}\right)^{-1} \ .
\end{equation}
The diffusion time scale decreases as a function of time. Therefore, it becomes shorter than the characteristic expansion time scale ($\sim R/V$) at some point, which roughly defines the time at the peak luminosity; 
\begin{equation}
t_{\rm peak} \sim 14 \ {\rm days}  \left(\frac{\kappa}{0.2 \ {\rm cm}^2 \ {\rm g}^{-1}}\right)^{1/2} \left(\frac{M_{\rm ej}}{M_\odot}\right)^{3/4}  \left(\frac{E}{10^{51} \ {\rm erg}}\right)^{-1/4} \ .
\end{equation}
This is, roughly speaking, the basic mechanism which determines the characteristic time scale in the rise and decay as seen in the light curves (Fig. \ref{fig:LC}). For example, if we take $M_{\rm ej} \sim 1.4 M_\odot$ and $E_{\rm K} \sim 1.3 \times 10^{51}$ erg, which is appropriate for SN Ia in the $M_{\rm Ch}$ WD scenario, we obtain $V \sim 7,000$ km s$^{-1}$ and $t_{\rm peak} \sim 17$ days. At the peak luminosity, the radius of the ejecta edge is $R \sim 10^{15}$ cm. In this phase, a large fraction of the ejecta must be opaque as is evident from the derivation above, so we can (roughly) approximate the SN emission as a blackbody with radius $R$. With the (observed) peak luminosity of $L \sim 10^{43}$ erg s$^{-1}$ (for SNe Ia), the characteristic temperature ($T$) can be estimated through the Stefan-Boltzmann law; 
\begin{equation}
T \sim 11,000 \ {\rm K} \left(\frac{L}{10^{43} \ {\rm erg} \  {\rm s}^{-1}}\right)^{1/4} \left(\frac{R}{10^{15} \ {\rm cm}}\right)^{-1/2} \ .
\end{equation}
Therefore, SNe emit most of the radiation power in the optical wavelengths. The spectral lines are formed above the photosphere, therefore in the outermost ejecta around the peak luminosity (see above). Therefore, the characteristic `spectral classification' (Fig. \ref{fig:classification}) reflects the composition in the outermost layer of the progenitor; type II for a progenitor with a H-rich envelope (e.g., an RSG), type Ib and Ic/Ic-BL for a He and C+O star, type Ia for a WD progenitor. 

As the spectral lines are formed within the expanding medium above the photosphere, the characteristic line profile is the P-Cygni profile \cite[e.g.,][]{branch2005,branch}, i.e., a combination of broad emission component and a blue-shifted absorption component (as seen in Fig. \ref{fig:spec}, except for SNe IIn and Ibn; see Section 5.3 for details) . The line width and the amount of the blueshift reflect the velocity at the line-forming region, typically slightly above the photosphere. Following the expansion and the density decrease, the photosphere is receding in mass coordinate as a function of time, therefore the photospheric velocity and line velocities usually decrease with time. 

\subsection{Power sources}

There are several options that can work as the sources of the SN luminosities. First option is the thermal energy deposited by the shock propagation through the progenitor star. From the equipartition, we expect that the thermal energy ($E_{\rm th}$) just after the shock breakout (i.e., the thermal energy at the initial radius of $R_0$ for the expanding ejecta, where $R_0$ is the progenitor radius) as follows; $E_{\rm th} (R_0) \sim E_{\rm K}$. In the optically thick phase, we may neglect the energy loss by radiation, then the thermal energy decreases as time goes by following the adiabatic expansion, i.e., $E_{\rm th} (R=R(t)) \sim E_{\rm K} (R(t)/R_0)^{-1}$ \cite{arnett}. We can then estimate the characteristic luminosity by computing the following; $L \sim E_{\rm th} (R=R(t_{\rm peak}))/t_{\rm peak}$. This expression reduces to the following; 
\begin{equation}
L \sim 10^{41} \ {\rm erg s}^{-1} \left(\frac{\kappa}{0.2 \ {\rm cm}^2 \ {\rm g}^{-1}}\right)^{-1} \left(\frac{R_0}{10^{11} \ {\rm cm}}\right) \left(\frac{E_{\rm K}}{10^{51} \ {\rm erg}}\right)  \left(\frac{M_{\rm ej}}{M_\odot}\right)^{-1} \ .
\end{equation}
Given that the typical (observed) luminosity of SNe is $\gsim 10^{42}$ erg s$^{-1}$, it is seen that this energy source is important for an RSG progenitor (SNe II) but not for a compact star (SNe Ia and Ib/c). 

Another important power source is the radioactive decay input, especially the decay chain of $^{56}$Ni $\to$ Co $\to$ Fe, with the e-folding time of 8.8 days and 111.3 days, respectively (see Chapter on `Radioactive decay'). In the earliest phase, the input power is approximately described as follow; 
\begin{equation}
    L (^{56}{\rm Ni}) \sim 6.5 \times 10^{43}  \exp\left(-\frac{t}{8.8 \ {\rm days}}\right) \frac{M (^{56}{\rm Ni})}{M_\odot} \ {\rm erg} \ {\rm s}^{-1} \ ,
\end{equation}
where we neglect the contribution from the $^{56}$Co decay and assume that all the $\gamma$-rays from the $^{56}$Ni decay are absorbed within the ejecta. The initial mass of $^{56}$Ni is denoted by $M$($^{56}$Ni). Later on, the contribution from the $^{56}$Co decay dominates the energy input. The escape of the $\gamma$-ray should also be taken into account toward the later phase; 
\begin{equation}
   L (^{56}{\rm Co}) \sim 1.5 \times 10^{43} \exp\left(-\frac{t}{111.3 \ {\rm days}}\right) \frac{M (^{56}{\rm Ni})}{M_\odot} \left(D_{\gamma} + f_{\rm e^{+}}\right)
   \ {\rm erg} \ {\rm s}^{-1} \ ,
\end{equation}
where $f_{\rm e^{+}} \sim 0.035$ is the fraction of the decay energy channeled to the positron emission (which is assumed to be locally thermalized within the ejecta). The $\gamma$-ray escape is described by $D_{\gamma}$, i.e., the fraction of the energy originally emitted as $\gamma$-rays but absorbed and thermalized within the ejecta; 
\begin{eqnarray}
    D_{\gamma} & \sim & 1-e^{-\tau_\gamma} \sim \tau_\gamma \ ({\rm for} \ \tau_\gamma \rightarrow 0)\nonumber\\
    \tau_\gamma &\sim & 1000 \frac{\left(M_{\rm ej}/M_\odot\right)^2}{\left(E_{\rm K}/10^{51} \ {\rm erg}\right)} \ \left(\frac{t}{\rm day}\right)^{-2}\ .
\end{eqnarray}
By setting $\tau_\gamma \sim 1$, we can derive the date ($t_\gamma$) when the $\gamma$-ray escape becomes substantial; 
\begin{equation}
    t_\gamma \sim 30 \ {\rm days} \left(\frac{M_{\rm ej}}{M_\odot}\right) \left(\frac{E_{\rm K}}{10^{51} \ {\rm erg}}\right)^{-1/2} \ .
\end{equation}
By comparing the peak time as determined by the optical-photon diffusion and the characteristic time scale for the $\gamma$-ray escape, it is seen that the full-trapping of the $\gamma$-ray power is a good approximation around the maximum-light phase. Then, the peak luminosity is roughly determined by the decay power at the peak date (usually dominated by the $^{56}$Co decay for most of SNe); this is $\sim 10^{41} - 10^{43}$ erg s$^{-1}$ for $M$($^{56}$Ni) $\sim 0.01 - 1 M_\odot$.

\begin{figure}[t]
\centering
\includegraphics[width=\columnwidth]{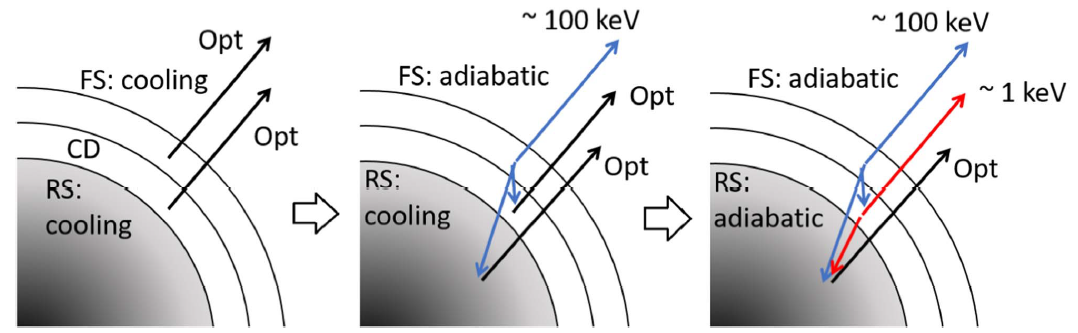}
\caption{Schematic picture of the SN-CSM interaction \cite{maeda2022}.
}
\label{fig:interaction}
\end{figure}

Yet another potential energy source is the ejecta-CSM interaction as shown schematically in Fig. \ref{fig:interaction}, for which a good review can be found in, e.g., \cite{chevalier2017}. As the ejecta are expanding into the CSM, two shock waves are formed \cite{chevalier1982a}; forward shock (FS) that defines the boundary between the shocked and unshoked CSM, and the reverse shock (RS) that defines the boundary between the shocked and unshocked ejecta. The shocked CSM and ejecta are connected by the contact discontinuity (CD). The dissipation rate of the kinetic energy at the shocks is estimated as follows; 
\begin{equation}
    L_{\rm int} \sim 2 \times 10^{42} 
    \left(\frac{\dot M / 0.01 M_\odot \ {\rm yr}^{-1}}{v_{\rm w} / 100 \ {\rm km} \ {\rm s}^{-1}}\right) 
    \left(\frac{V_{\rm SN}}{4,000 \ {\rm km} \ {\rm s}^{-1}}\right)^3
    \ {\rm erg} \ {\rm s}^{-1} \ .
\end{equation}
Here, $V_{\rm SN}$ is the expansion velocity of the CD. As we see, the interaction power can dominate over the other energy sources (the initial thermal energy and/or the radioactive energy input) for an  extremely high mass-loss rate of $\gsim 0.01 M_\odot$ yr$^{-1}$ (assuming the mass-loss wind velocity of $v_{\rm w} \sim 100$ km s$^{-1}$). We note that the materials passing through the forward and reverse shocks are initially heated to temperature emitting X-ray photons \cite{chevalier1982b,chevalier2006,chevalier2017,maeda2022}. For example, for the FS, the temperature will be $\sim 10^9 (V_{\rm SN}/10,000 \ {\rm km} \ {\rm s}^{-1})^2$ K (see Chapter on `Thermal processes in supernova remnants'), with the corresponding photon energy of $\sim 100$ keV. The temperature behind the RS is lower by one or two orders of magnitude, depending on the ejecta and CSM structures, leading to the corresponding photons energy of $\sim$ a few keV. 
 
The evolution of $V_{\rm SN}$ depends on the properties of the ejecta and the CSM; if the density slopes of the ejecta and the CSM are both described by a single power-law, $n$ for the ejecta and $s$ for the CSM (i.e., $\rho_{\rm ej} \propto v_{\rm ej}^{-n}$ and $\rho_{\rm CSM} \propto r^{-s}$, noting that the unshocked ejecta follow the homologous expansion, $v_{\rm ej} \propto r/t$), it can be derived that $V_{\rm SN}$ evolves as follows \cite{chevalier1982a,moriya2013}, from the dimensional analysis; 
\begin{equation}
V_{\rm SN} \propto t^\frac{s-3}{n-s} \ .
\end{equation}
The value of $n$ is described as a steep power law with $n \sim 7 -13$ for typical models, with a compact progenitor and extended progenitor result is shallow and steep gradient \cite{matzner1999}. For a steady-state mass-loss wind with a constant velocity, $s=2$. If we take $n=10$ and $s=2$, $V_{\rm SN} \propto t^{-0.125}$, and thus $L_{\rm int} \sim t^{-0.375}$. As a specific example, the following expression has been frequently applied for an explosion of a He or C+O progenitor, with $n=10.18$ \cite{chevalier2006}; 
\begin{equation}
V_{\rm SN} \sim 8  \times 10^{9} \ {\rm cm} \ {\rm s}^{-1} \left(\frac{E_{\rm K}}{10^{51} \ {\rm erg}}\right)^{0.43} \left(\frac{M_{\rm ej}}{M_\odot}\right)^{-0.32} A_{*}^{-0.12}
\left(\frac{t}{{\rm day}}\right)^{-0.12} \ , 
\end{equation}
where $A_{*}$ is a normalization constant for the CSM density as defined by $\rho_{\rm CSM} = 5 \times 10^{11} A_{*} r^{-2}$ g cm$^{-3}$ where $r$ is in cm. Namely, 
\begin{equation}
A_{*} \sim \left(\frac{\dot M}{10^{-5} M_\odot {\rm yr}^{-1}}\right)
\left(\frac{{v_{\rm w}}}{1,000 {\rm km} \ {\rm s}^{-1}}\right)^{-1} \ , 
\end{equation}
and $A_{*} \sim 10^4$ for $\dot M \sim 0.01 M_\odot$ yr$^{-1}$ and $v_{\rm w} \sim 100$ km s$^{-1}$. We then have the following for the interaction power; 
\begin{equation}
L_{\rm int} \sim 10^{43} \ {\rm erg} \ {\rm s}^{-1} 
 \left(\frac{E_{\rm K}}{10^{51} \ {\rm erg}}\right)^{1.29} \left(\frac{M_{\rm ej}}{10 M_\odot}\right)^{-0.96} \left(\frac{A_{*}}{10^4}\right)^{0.64} \left(\frac{t}{100 \ {\rm days}}\right)^{-0.36} \ .
\end{equation}

\subsection{SN Progenitors and Explosions as Seen in Observations}

With the basic properties of the SN emission as introduced in the previous sections, it is possible to connect different types of progenitors and explosion mechanisms to SN observables. In addition, there are other observational constraints placed on the nature of the progenitor and explosion, including the nature of progenitor candidates detected in pre-SN images \cite[e.g.,][]{smartt2009}. 

{\bf Type Ia SNe: }
As mentioned in Sections 1 and 5.1, spectral properties of SNe Ia suggest a relation to a WD thermonuclear explosion. The typical velocity seen in spectral lines, e.g., Si II 6355, is $\sim 10,000$ km s$^{-1}$, roughly consistent with the expected ejecta velocity for the WD thermonuclear explosions ($M_{\rm ej} \sim 1-1.4 M_\odot$ and $E_{\rm K} \sim (1-1.5) \times 10^{51}$ erg) (note that the `average' velocity estimated by equation 10 is underestimate for the spectral velocity in the `outermost' layer seen around the maximum light). The characteristic time scale in the light curve evolution is $\sim 15-20$ days (depending on the band passes), which is again consistent with the range of $M_{\rm ej}$ and $E_{\rm K}$ expected in the WD thermonuclear explosions. 

The light curve evolution is well explained by the $^{56}$Ni/Co/Fe decays as a main power source. The peak luminosity is $\sim (1-2) \times 10^{43}$ erg s$^{-1}$, i.e., $M$($^{56}$Ni) $\sim 0.5-1 M_\odot$. This is a range well explained by the delayed-detonation model and the double-detonation model (Sections 3 and 4). 

For a typical WD radius ($\sim 10^9$ cm), the thermal energy stored during the explosion will never be a dominant power as it is quickly gone as a result of the adiabatic cooling (eq. 15). This indeed has lad to several diagnostics proposed to discriminate between the SD and DD progenitor scenarios (e.g., \cite{maeda2016} for a review). An important difference in the SD scenario from the DD is the existence of a non-degenerate companion. The collision between the SN ejecta and a non-degenerate companion (e.g., \cite{liu2013}) can generate and store the thermal energy \cite{kasen2010a}; for an order-of-magnitude estimate, one can insert the binary separation (which is comparable to the radius of the companion, as it is filling the RL; $\sim 10^{11} - 10^{13}$ cm) into equation 15. The expected luminosity can overwhelm the $^{56}$Ni power in the first few days, and it is a level that is detectable by recent high-cadence surveys. This is a topical issue with an increasing number of samples available for such investigation (i.e., those discovered soon after the explosion and then promptly and intensively followed-up) in the last few years. The latest observational results can be found in, e.g., \cite{burke2022}. Indeed, various methods have been proposed to discriminate between different progenitor scenarios (e.g., SD vs. DD) as well as different explosion mechanisms (e.g., delayed detonation vs. double detonation), including a search for a (surviving) companion star within SNRs in the MW and LMC as well as in pre-and post-SN images of nearby SNe. Covering all these topics is beyond the scope of this contribution, and we refer the readers to \cite{maeda2016} for a review.


Last but not least, it must be emphasized that SNe Ia are a mixture of various subclasses and outliers \cite{taubenberger2017}. While their general observational features point to the thermonuclear WD explosion origin, the details differ. It has thus been suggested that SNe Ia may indeed originate through several evolution channels and explosion mechanisms. For example, there are a few peculiar SNe Ia for which the double detonation is a leading interpretation \cite{jiang2017,de2019}. There is also a class of peculiar SNe Ia for which an explosion of a $M_{\rm Ch}$ C+O WD, but only by deflagration without the detonation, is a leading scenario \cite{jha2017}. It has not been clarified what evolutionary channel and explosion mechanism are responsible to the main population of SNe Ia \cite[e.g.,][for a review]{maeda2016}

{\bf Type IIP/L SNe: } The expected RSG progenitors, with its large luminosity in the optical, make it practically possible to search for a progenitor star in pre-SN images mainly provided by Hubble Space telescope archival data; it is called `the direct progenitor detection'. It is a challenging observation, but the progenitor candidates have been routinely detected for a handful of very nearby SNe IIP \cite[e.g.,][]{smartt2009,smartt2015}. In many cases, the expected RSG progenitors are recovered, and thus the RSG progenitors for SNe IIP have been solidly established. From statistical analyses, the range in $M_{\rm ZAMS}$ leading to SNe IIP as the final outcome has been derived to be $\sim 9 - 18 M_\odot$. Within uncertainties, it shows a good match to the theoretical expectation from the standard stellar evolution models (Section 2), while the relatively low value for the upper limit is puzzling; while we have RSGs in the MW with $M_{\rm ZAMS} \gsim 18 M_\odot$, such stars do not explode as SNe IIP, against the standard expectation ($\sim 25-30 M_\odot$ depending on details of the stellar evolution model; e.g., Fig. \ref{fig:mass}). This is called the RSG problem \cite{smartt2009}. 

The strong and long-lasting Balmer lines (with the P-cygni profile), as seen in Fig. \ref{fig:spec}, indicate that the photoshpere keeps being formed within a massive H-rich envelope, again suggesting an RSG progenitor. The typical mass of the H-rich envelope is $\sim 10 M_\odot$ (Fig.\ref{fig:mass}), and we may assume $M_{\rm ej} \sim 10 M_\odot$ together with $E_{\rm K} \sim 10^{51}$ erg. We then estimate that the typical timescale is $\sim 100$ days, which matches to the `plateau' seen in SNe IIP (Fig. \ref{fig:LC}). While SNe IIL do not show the pleateau as clearly as in SNe IIP, it is seen that the characteristic timescale is similar (or a bit shorter) (Fig. \ref{fig:LC}). The velocity in the spectral lines is smaller for SNe IIP/IIL than SNe Ia and II/Ib/Ic (see below), which is also in line with the small ratio of $E_{\rm K}/M_{\rm ej}$. 

We however note that this is a very simplified picture. For example, the opacity is never constant in space and time; the recombination of H, which drastically changes the opacity across the recombination front, is indeed a key in determining the LC properties of SNe IIP/L given its massive H-rich envelope, and this must be taken into account in comparing the model and data \cite{arnett}. In any case, the RSG progenitor has been shown to be largely consistent with observational properties of SNe IIP based on numerical radiation transfer simulations \cite[e.g.,][]{dessart2013}. The origin of SNe IIL is less clear, with several possible interpretations for its rapidly-declining LC evolution, e.g., it could be due to a relatively small amount of the H-rich envelope \cite{moriya2016} and/or the substantial contribution from the SN-CSM interaction \cite{morozova2017}. 

SNe IIP/L experience the initial `cooling phase' with the time scale of 10-20 days, before the onset of the recombination within the ejecta. This phase can also be modeled with a similar idea for the plateau phase but with the temperature evolution mainly determined by the adiabatic cooling rather than the recombination \cite{rabinak2011}. This is a UV-birght emission given its high temperature, with rising optical LCs following the blackbody peak moving into the optical range due to the temperature decrease. This is a powerful diagnostics to derive the progenitor radius (see eq. 15), and has been applied to a sample of SNe IIP/L. 

However, the increasing sample has shown discrepancy between the observed multi-band LCs in the first 10-20 days and the model expectation based on the standard stellar evolution models \cite{morozova2015,forster2019}; the standard progenitor models predict a monotonically-rising multi-band optical LCs in this initial phase which then merge smoothly into the plateau phase, while the observed SNe IIP/L typically show a bump in the optical. A popular idea to explain this behavior is an existence of a confined and dense CSM within $\sim 10^{15}$ cm. The corresponding mass-loss rate in the last few decades (for $v_{\rm} \sim 10$ km s$^{-1}$) can reach to $\sim 10^{-3} - 0.1 M_\odot$ yr$^{-1}$, far exceeding a conventional mass-loss rate adopted in standard stellar evolution models. This is further supported by the `flash spectra' \cite{gal-yam2014,yaron2017}; a good fraction of SNe IIP/L show narrow (unresolved, $\lsim 500$ km s$^{-1}$) emission lines of highly-ionized ions superimposed on a blue continuum in infant spectra taken within a day or a few days since the explosion, which is totally different from typical SN IIP (and IIL) spectra with Balmer series seen as a broad ($\sim 10,000$ km s$^{-1}$) P-cygni profile (e.g., Fig. \ref{fig:spec}). This flash spectrum is interpreted as an outcome of initial SN radiation producing UV ionizing photons (i.e., blue continuum) that passes through a dense CSM. This is  followed by a quick recombination within the CSM (i.e., narrow emission lines from highly-ionized ions). Its short duration limits the size of the dense CSM as $\sim 10^{15}$ cm (i.e., the light-travel time for the flash spectrum lasting only for $\sim 10$ hrs). 

The discovery of the confined and dense CSM has led to a paradigm change in the stellar evolution study. The origin has not yet been clarified, but it is likely related to the rapid evolution of the stellar core in the final phase; some key processes in this `final phase' are probably still missing in our knowledge of the stellar evolution. This discovery has been driving further intensive study both by observational approaches (through infant SNe; \cite[e.g.,][]{forster2019,maeda2021}) and theoretical approaches (through the interpretation of the infant SN observational data and stellar evolution simulations; \cite[e.g.,][]{fuller2017,ouchi2009}). 

{\bf Type IIb/Ib/Ic SNe: }
The direct progenitor detection for SNe Ib and Ic has been very limited, with mostly upper limits derived for nearby objects \cite{smartt2009}. The upper limits sometimes go deeper than the expected magnitudes of Galactic WR stars, indicating that at least a fraction of SNe Ib/c progenitors are different from the Galactic WR population. There is one strong progenitor candidate detected for SN Ib (iPTF13bvn) \cite{cao2013,folatelli2016}: its properties favor the binary evolution scenario but a massive single star scenario has not been completely rejected. A key challenge is that the expected bare He or C+O progenitors are luminous in the UV, but relatively faint in the optical in which most of the pre-SN data are available. A recent detection of bright and red (i.e., extended) point source for SN Ib 2019yvr \cite{kilpatrick2021} adds further complication to this picture, which might indicate existence of a dense CSM around some SNe Ib, perhaps similar to the case for (some) SN IIP progenitors (see above). 

The direct progenitor investigation has turned out to be successful for SNe IIb \cite[see][for a review]{smartt2015}. Interestingly, the progenitors' (or candidates') radii show a large diversity, from a blue-supergiant (BSG) size ($\sim 50 R_\odot$) to an RSG ($\gsim 500 R_\odot$), including a yellow-supergiant (YSG) dimension ($\sim 200 R_\odot$). This is interpreted to reflect the amount of the remained H-rich envelope, as the hydrostatic structure changes at the boundary of $M_{\rm H} \sim 0.1 M_\odot$; the BSG for $M_{\rm H} \lsim 0.1 M_\odot$, YSG for $M_{\rm H} \sim 0.1 M_\odot$, and RSG for $M_{\rm H} \gsim 0.1-1 M_\odot$. For a more massive H-rich envelope, the SN will be observed as either SNe IIL or IIP. This sequence is consistent with the binary evolution scenario, in which the initial separation controls the final mass of the H-rich envelope and the radius of the progenitor \cite{ouchi2017}. 

The SN properties (light curves and spectra) of SNe IIb/Ib/Ic favor a relatively low-mass He or C+O star (with no or thin H-rich envelope), over a massive WR progenitor expected in the single stellar evolution. The characteristic time scale and the spectral line velocity are comparable to those of SNe Ia, and the typical ejected mass has been estimated to be $M_{\rm ej} \sim 1 - 3 M_\odot$ \cite{lyman2016}. As shown in Fig. \ref{fig:mass}, it corresponds to the mass range of $M_{\rm ZAMS} \sim 10-20 M_\odot$, as is similar to SNe IIP. As such, it indicates that the major evolution channel toward SNe IIb/Ib/Ic is a binary evolution; they would become SNe IIP if they would not be in a close binary system. 

The bright initial cooling phase is expected for (some) SNe IIb with extended progenitors, with a typical time scale of a few days to a week, i.e., shorter than SNe IIP due to the smaller amount of the H-rich envelope. It has been used to infer the progenitor radius for the SN IIb progenitors, resulting in generally a consistent result with the direct progenitor detection \cite[e.g.,][]{bersten2012}. The similar observational investigation is challenging for SNe Ib/Ic for their smaller radius, but the increasing examples of infant SN observations start producing meaningful upper limit for the radius of their progenitors \cite[e.g.,][]{stritzinger2020}.  

{\bf Type IIn and Ibn/Icn SNe: }
There are SNe showing signatures of strong interaction between the SN ejecta and dense CSM. The population of SNe IIn is a classical example, as characterized by strong Balmer series in emission lines (Fig. \ref{fig:spec}).  When a high-dispersion spectroscopy is performed, a narrow component ($\sim 100$ km s$^{-1}$) is frequently seen in the P-cygni profile overlapped on a broader component; it is interpreted that the narrow component originates in the unshoked dense CSM, while the broad component represents either the shocked region or unshocked ejecta \cite[e.g.,][for a review]{smith2017}. 

Their LCs show diversity, with bright ones exceeding the peak luminosity of SNe Ia (Fig. \ref{fig:LC}). These bright SNe IIn typically show slowly-evolving LCs; the total energy in the radiation thus can exceed $10^{50}$ erg, sometimes reaching to $\sim 10^{51}$ erg. This huge energy output, together with the spectral properties, suggests that they are fully powered by the SN-CSM interaction, where the energy budget is attributed to the kinetic energy of the ejecta dissipated through the interaction. It is seen in equation 24 that the luminosity of $\gsim 10^{43}$ erg s$^{-1}$ can be explained by the CSM corresponding to the pre-SN mass-loss rate of $\gsim 0.01 M_\odot$. This is confirmed by analyses of SN IIn LC in a more detailed manner \cite[][]{moriya2013}, with the estimated mass-loss rate in the range of $\sim 10^{-3} - 1 M_\odot$. A caveat here is that such analysis might be biased toward bright objects (with high mass-loss rates). 

The progenitors of SNe IIn have not been clarified. With the huge mass budget, it is generally believed to be a massive star, perhaps those even more massive than other classes of CCSNe (see \cite{gal-yam2009} for a possible progenitor detection). However, the mechanism and origin of the huge mass-loss rate have not yet been understood, which might be either a Luminous-blue variable like eruption, RL mass transfer in a binary, or even involve a CE evolution. 

SNe Ibn are a He-rich analog of SNe IIn. Instead of Balmer lines, a number of He lines are seen as emission lines (Fig.\ref{fig:spec}), pointing to the He-rich CSM \cite[e.g.,][]{pastorello2007}. Therefore the ejecta should also be H-poor, and thus they could also be considered as an analog of SNe Ib or Ic but with dense (He-rich) CSM. Interestingly, SNe Ibn show different characteristics from SNe IIn in their LC evolution \cite{hosseinzadeh2017,maeda2022} (Fig. \ref{fig:LC}); they typically evolve much faster than SNe IIn, which indicates a rapidly increasing mass-loss rate toward the SN \cite{maeda2022}. The progenitor origin of SNe Ibn, as well as its relation to SNe IIb/Ib/In, have not been clarified, but it is likely that the progenitors of SNe Ibn are intrinsically different from those of SNe IIb/Ib/In; for example, an idea has been proposed to associate SNe Ibn to a massive WR star (essentially through a single star evolution) while a majority of SNe IIb/Ib/Ic are an outcome of a binary evolution of less massive stars \cite[e.g.,][]{pastorello2007,maeda2022}. Very recently, a carbon-rich analog of SNe Ibn has been discovered, coined SNe Icn \cite{gal-yam2022}; this newly-found population will provide key information on the progenitor evolution of strongly interacting SNe. 

As a variant of SNe IIn showing signatures of strong SN-CSM interaction, the so-called SNe Ia-CSM have been suggested to be associated with a thermonuclear WD explosion rather than CCSNe; they show similarities with SNe Ia in their broad spectral features \cite{hamuy2003}, and there is one example which showed clear transition from SN Ia to SN Ia-CSM with increasing strength of SN-CSM interaction in the late phase \cite{dilday2012}. The origin of SNe Ia-CSM has not been clarified, with a major challenge in explaining how a large amount of CSM could be produced in relative low-mass systems involving a WD. At least for the most extreme cases, it may involve a CE phase to produce the dense CSM environment, and the thermonuclear WD explosion might be triggered even as an outcome of the core merger through the CE \cite{jerkstrand2020}.

{\bf Other variants: }
Properties of SNe are indeed very diverse, and thus this review can cover only a fraction of SN subclasses. Here, we briefly mention on some peculiar and interesting subclasses; the readers are refereed to \cite{branch} as a comprehensive review on this topic. 

\begin{figure}[t]
\centering
\includegraphics[width=0.55\columnwidth]{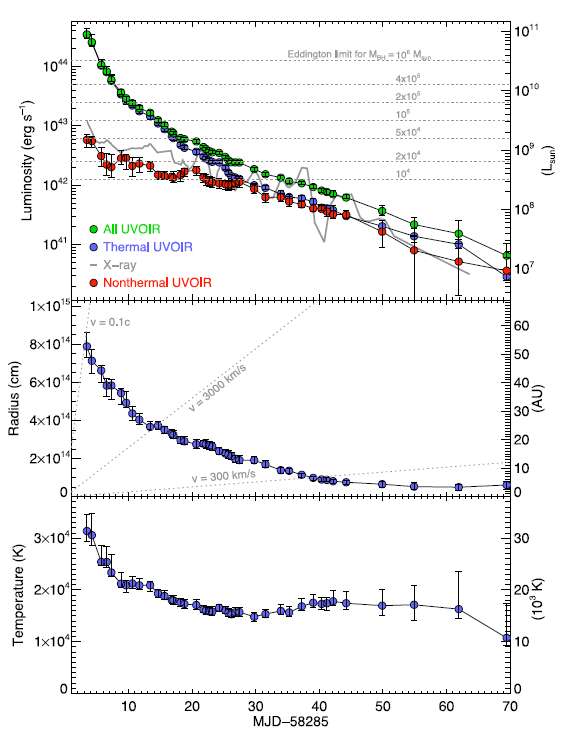}
\includegraphics[width=0.41\columnwidth]{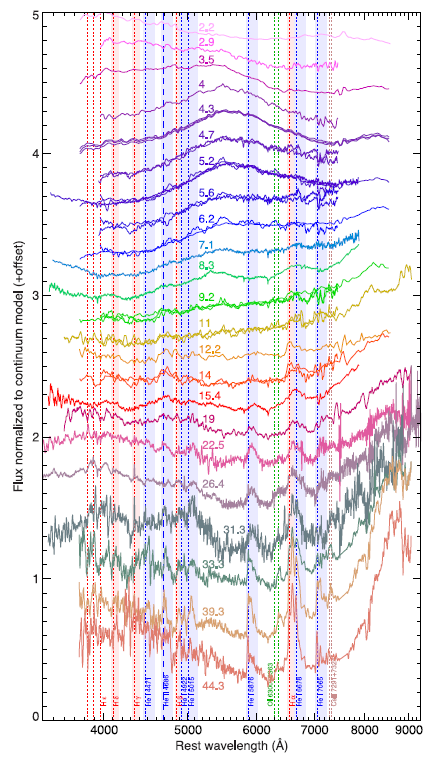}
\caption{Observational properties of AT 2018cow \cite{perley2019}. (Left: ) Evolution of the bolometric luminosity, photospheric radius and temperature as derived by a blackbody fit. (Right: ) Spectral evolution. 
}
\label{fig:2018cow}
\end{figure}

In relation to SNe Ic, there is a class of SNe Ic but with much broader spectral lines in their spectra, called SNe Ic-BL (Fig. \ref{fig:spec}). The prototypical SN Ic-BL is SN 1998bw, which was discovered in association with a long/soft Gamma-Ray Burst (GRB) 980425 \cite{galama1998}. To explain a combination of broader spectral features (i.e., higher ejecta velocity) and longer time scale in its LC (i.e., longer diffusion time scale) than canonical SNe Ic, it has been argued that SN 1998bw has massive ejecta ($\sim 10 M_\odot$) and a large kinetic energy ($E_{\rm K} \gsim 10^{52}$ erg) \cite{iwamoto1998}. A sample of SNe Ic-BL, some are associated with a GRB (GRB-SNe) but others are not, have been discovered, forming a class of SNe Ic-BL \cite[][for a review]{woosley2006}. It turns out that GRB-SNe (including SN 1998bw) indeed represent the most extreme example in its ejecta mass and kinetic energy. These properties seem to be less extreme for SNe Ic-BL without associated GRBs, but still they are generally more energetic than canonical SNe Ic. The energy budget required for SNe Ic-BL (especially for GRB-SNe) is far beyond the expectation within the framework of the standard delayed neutrino explosion scenario for CCSNe (Section 2.2), and thus different mechanisms have been proposed; two popular ideas are (1) formation of a black hole (BH) and an accretion disk inside a collapsing massive star \cite{macfadyen1999}, which energize both relativistic jets (to explain GRBs) and energetic SN ejecta (to explain the SN component), and (2) formation of a highly magnetized NS (an analog of a magnetar but with rapid rotation) \cite{zhang2001,metzger2011}. Both scenarios involve an efficient conversion of the gravitational binding energy to the final outflow energy, by storing the energy first in form of the rotational energy; the conversion of the rotation energy to the outflow energy is likely mediated by magnetic field. The progenitors have been suggested to be a rapidly-rotating massive star \cite[e.g.,][]{paczynski1998}, the formation of which will require a special evolutionary channel either in a single or binary scenario \cite{yoon2005}. 

Another class of objects of interest is a population of super-luminous SNe (SLSNe) \cite[e.g.,][for a review]{gal-yam2012}. As in the classical classification scheme, they are divided into type II (H-rich; SLSNe-II) and I (H-poor; SLSNe-I). SLSNe-II are observationally a (even more) luminous variant of SNe IIn, and there is little doubt that they are mainly powered by the strong SN-CSM interaction. The required CSM is larger than SNe IIn, and the nature of the progenitor system has not been clarified yet. The nature of SLSNe-I is less clear; even the power source to keep thier high luminosities has not yet been robustly identified. A popular suggestion is that they might be powered by the spin-down energy of highly-magnetized NS (or an actively accreting BH) in the center,  produced by the core collapse and explosion \cite{maeda2007,kasen2010b}. The central engine therefore may be similar to the one for GRB-SNe, and they may be unified into a single scheme \cite{greiner2015,nicholl2016,suzuki2021}. 

Rapidly-evolving transients form a new class of transients that have been discovered recently thanks to the new-generation, wide-field and high-cadence surveys \cite[e.g.,][]{drout2014}. Their short time scale, with characteristic time scale of $\lsim$ 10 days, has been a hurdle for the real-time prompt follow-up observations, and thus the detailed information such as spectra had been largely missing until recently. The first systematic sample with the spectral classification has been recently reported for those discovered by Zwicky Transient Facility (ZTF) \cite{ho2021}. It shows that the rapidly-evolving transients are composed of different populations, mainly divided into three classes; relatively low-luminous SNe IIb/Ib, luminous SNe Ibn/IIn, and extremely rapid and luminous (and rare) AT 2018cow-like objects. AT 2018cow-like objects are very different from known SN populations in various observational properties \cite{perley2019} (Fig. \ref{fig:2018cow}); they are extremely rapid, raising to the luminous peak ($\lsim -22$ mag, as is comparable to SLSNe) at most in 5 days (or even less). The color in the UV/optical does not evolve much, unlike canonical SNe showing the blue-to-red evolution. The optical spectra are also different from SNe, with some similarity to tidal-disruption events. Interestingly, there is indication of a powerful central engine as seen in X-rays from AT 2018cow \cite{margutti2019}, which might also be responsible to its special appearance in the UV/optical/NIR wavelengths. No consensus has been reached to its origin and even the power source, with various scenarios suggested so far, including, e.g., a BH formation following a failed SN explosion of a massive star (as is similar to a model for a long GRB), a pulsational pair-instability SN originated in a very massive star, an electron-capture SN, a tidal disruption event by a super-massive BH, an explosion following a CE, among others \cite[e.g.,][and references therein]{uno2020}. 

\subsection{High-Energy Emissions from Supernovae}

\begin{figure}[t]
\centering
\includegraphics[width=\columnwidth]{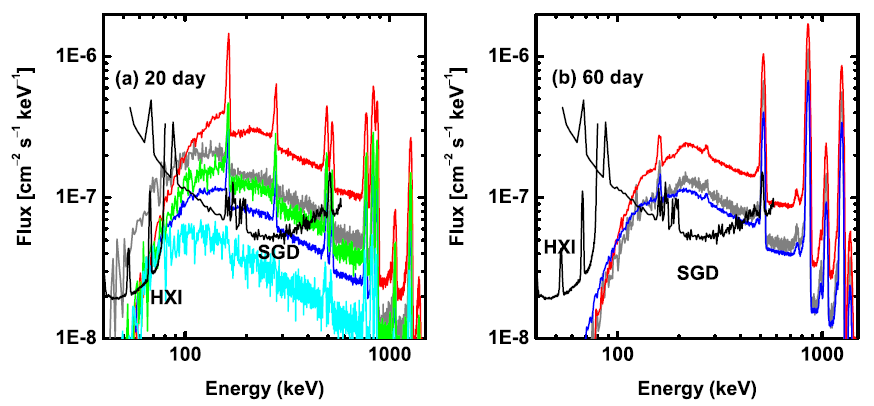}
\caption{Examples of synthetic spectra for hard-X and $\gamma$-rays as a result of radioactive decay chain $^{56}$Ni $\to$ $^{56}$Co $\to$ $^{56}$Fe for a few SN Ia models \cite{maeda2012a}. 
}
\label{fig:radioactivity}
\end{figure}

There are several mechanisms that are responsible to high-energy emissions from (young) SNe. We will list some of them in this section. 

Emission following decays of radioactive isotopes is one mechanism (see Chapter on `Radioactive decay' for details). In SNe, the most abundantly produced unstable isotope is $^{56}$Ni (Section 4), which is also responsible to optical emissions (Sections 5.2 and 5.3). Examples of the simulated high-energy emissions in the hard-X and MeV-$\gamma$ rays are shown in Fig. \ref{fig:radioactivity} for a few SN Ia models. The decay chain $^{56}$Ni $\to$ $^{56}$Co $\to$ $^{56}$Fe produces characteristic MeV $\gamma$-ray lines, which interact with gas in the expanding SN ejecta primary through Compton scattering. The degraded $\gamma$-rays through the Compton scattering create a continuum down to $\sim 100$ keV below which the photons are absorbed through the photoelectric absorptions. Fig. \ref{fig:radioactivity} shows strong lines from the $^{56}$Ni decay on day 20 (e.g., 128 and 812 keV) and from the $^{56}$Co decay on day 60 (e.g., 847 keV). As time goes by, a fraction of the escaping lines increases (see equations 18 and 19) while the decay power decreases exponentially (equations 16 and 17); the combination creates the peak in the flux evolution of the decay lines, e.g., $\sim 60-80$ days for the 847 keV line in the case of SNe Ia. 

These signals have been detected in SN II 1987A and SN Ia 2014J. For SN 1987A, the radioactive decay emission from the $^{44}$Ti $\to$ $^{44}$Sc $\to$ $^{44}$Ca has additionally been robustly detected. Another example of the $^{44}$Ti/Sc/Ca decay emission is the one from Cas A SN remnant. Chapter on `Radioactive decay' provides a comprehensive description on these observations (see the descriptions and references therein). Given that synthesis of radioactive isotopes traces the innermost part of the exploding core (Section 4), these radioactive decay emissions serve as an irreplaceable probe to the explosion mechanism and the core properties of the progenitor stars.

\begin{figure}[t]
\centering
\includegraphics[width=\columnwidth]{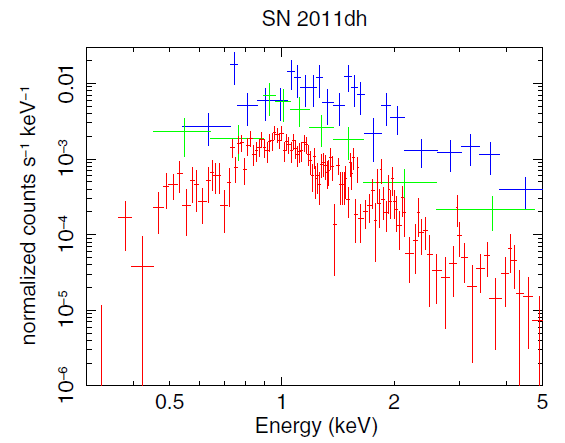}
\caption{An example of X-ray emission \cite{maeda2014}; it is for SN IIb 2011dh on day 12 (blue), 33 (green), and $\sim 500$ (red). The data shown here have been taken by the Chandra observatory. 
}
\label{fig:x_2011dh}
\end{figure}

The SN-CSM interaction is another major emission process of high-energy photons from young SNe. As this originates in the FS and RS, this signal is an important probe to the natures of the CSM and the outermost layer of the progenitor star \cite[][for a review]{chevalier2017}. If the shocks are in the adiabatic regime (i.e., with negligible cooling), the characteristic temperature is $\sim 100$ keV at the FS and $\sim 1$ keV at the RS; the most of the energy is thus emitted as thermal hard-X and soft-X photons, for the FS and RS, respectively. However, the RS can easily be in the cooling regime and thus does not produce high-energy photons unless the CSM density is extremely low, and thus the X-ray emission is initially dominated by the one from the FS, essentially through free-free emission. Later on, the RS becomes adiabatic, and thermal emission from the RS can dominate in the soft X-ray photons. An additional mechanism is the inverse Compton (IC) scattering of the thermal optical photons from the ejecta by relativistic, non-thermal electrons produced at the FS. Further consideration must be provided for the absorption process; especially important is the photo-electric absorption in the soft X-ray band, mainly through the unshocked CSM (plus some contribution in the shocked FS and RS regions), the effect of which is highly dependent on the CSM density and composition. 

Fig. \ref{fig:x_2011dh} shows the evolution of the X-ray spectra for SN IIb 2011dh, given as a specific example. Its thermal nature at $\sim 500$ days as originated in the RS is robustly identified \cite{maeda2014}. The data at day 33 are also fit well by the RS thermal emission model. The X-ray emission on day 10 is probably contributed substantially by the FS component, either through the free-free emission or the IC \cite{sasaki2012}. By identifying the thermal component, it is possible to robustly derive the CSM density, therefore the mass-loss rate. The result can further be combined with the radio synchrotron emission to infer the nature of the acceleration of relativistic, non-thermal electrons at the FS \cite{maeda2012b}; the radio synchrotron model alone suffers from the degeneracy in the nature of the particle acceleration and the CSM density, but by adding the X-ray data this can be at least partly solved. 

\begin{figure}[t]
\centering
\includegraphics[width=\columnwidth]{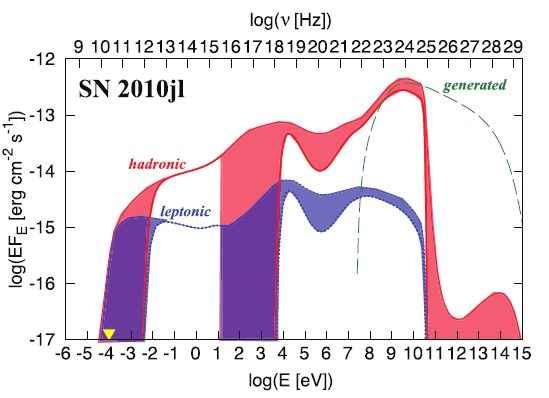}
\caption{An example of synthetic multi-wavelength emissions from an SN IIn, powered by a strong SN-CSM interaction \cite{murase2019}. 
}
\label{fig:multi_spec}
\end{figure}

For SNe with extremely dense CSM, emissions from relativistic, non-thermal protons/ions accelerated at the FS can produce strong contributions across various wavelengths. Fig. \ref{fig:multi_spec} shows such a model example, for SN IIn 2010jl; this is a very bright SN IIn and its CSM density is in the highest regime even for SNe IIn. The non-thermal protons produce $\gamma$-rays at the Gev and TeV ranges through the pion decay following the interaction with thermal protons. Secondary electrons thus produced can play an important role in the radio synchrotron emission together with the primary electrons directly accelerated at the FS. Even neutrinos at Gev and TeV ranges are produced; SNe with very strong SN-CSM interaction have thus been proposed as one of possible electromanetic counterparts for TeV neutrinos (e.g., those detected at the IceCube), i.e., a target in the multi-messenger astronomy \cite[e.g.,][]{murase2019}. 

\begin{figure}[t]
\centering
\includegraphics[width=\columnwidth]{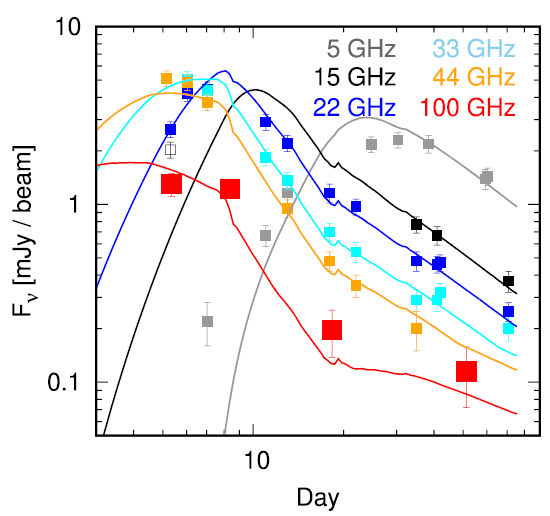}
\caption{An example of radio emission, including the mm emission observed by the ALMA \cite{maeda2021} and the cm emission from VLA and other facilities \cite{horesh2020}. SN Ic 2021oi shown here is a rare example for which multi-band light curves including the mm wavelengths are obtained within a week after the explosion. 
}
\label{fig:radio}
\end{figure}

As mentioned above, the radio emission from SNe originates in the electrons accelerated at the FS in most cases (except for the extremely dense CSM in which the secondary electrons' contribution becomes substantial; Fig. \ref{fig:multi_spec}). As such, the observation and analysis of radio signals provide a powerful probe to the nature of the CSM \cite[][for a review]{chevalier2017}. One of the recent highlights is investigation of the nature of the CSM in the very vicinity of the SN progenitor. Thanks to the new-generation (optical) transient surveys that routinely discover SNe soon after the explosion (within a few days), radio observations of infant SNe (within a week of the explosion) are now practically possible, which provide the nature of the CSM within $\sim 10^{15}$ cm. This is a powerful method to probe the nature of the mass loss just before the explosion; $10^{15}$ cm $/ v_{\rm w} \sim 30$ yrs for the RSG progenitor ($v_{\rm w} \sim 10$ km s$^{-1}$; SNe IIP) or even $\sim 0.3$ yrs for the compact He or C+O progenitors ($v_{\rm w} \sim 1,000$ km s$^{-1}$; SNe Ib/c). Given the high CSM density toward the inner region even for a constant mass-loss rate and wind velocity, high-frequency (mm) observation is a key thanks to its transparency \cite{matsuoka2019}. 

Fig. \ref{fig:radio} shows such an example for SN Ic 2020oi; assuming $v_{\rm w} \sim 1,000$ km s$^{-1}$, the analysis of the data allows to trace the mass-loss history down to the final $\sim 0.5$ yr before the explosion. The ALMA data trace the optically-thin regime, and the observed temporal evolution, the flat-steep-flat evolution, is interpreted as the CSM density having the same, flat-steep-flat distribution from the inner to the outer regions, i.e., it is not described by a steady-state mass-loss mechanism. The fluctuation seen for the final mass-loss properties in the sub-year time scale indicates that the origin of the final activity is related to the accelerated change in the core nuclear burning stage, and that the envelope reacts to the core evolution dynamically; this finding further strengthens the argument for existence of the (yet-unclarified) final activity of (at least some of) massive stars in the final decades as has been probed by the optical emission (Section 5.3), and pushes the investigation further down to the final months before the SN explosion.

\section{Summary}
The fields of stellar evolution and SN explosion, as well as the transient observations (including SNe), are quickly developing with inflating opportunities both in theoretical and observational approaches. In this chapter, we have provided an overview on the stellar evolution channels toward SNe, explosion mechanisms, and explosive nucleosynthesis. There are many unresolved issues in these fundamental processes in astronomy, and studying properties of SNe is a unique and powerful probe to study these issues which are otherwise difficult. Observational properties of SNe, covering the basic and classical pictures, as well as up-to-date recent findings, are also summarized in this chapter, with which we demonstrate how the natures of the progenitor stars and the explosion mechanisms can be constrained. The contents of this chapter aim at providing usual basis and guideline in this quickly-developing field, and the readers are encouraged to access review articles and recent papers on individual subjects introduced in this chapter.

\section{Acknowledgement}
K.M. thanks Avinash Singh, Anjasha Gangopadhyay, and Kohki Uno for their assistance to produce Figs. \ref{fig:LC} and \ref{fig:spec}. K.M. acknowledges support from the Japan Society for the Promotion of Science (JSPS) KAKENHI grant JP18H05223 and JP20H00174. Some data presented in this contribution are obtained from WISeREP (https://www.wiserep.org).




\end{document}